\documentclass[11pt,a4paper]{article}
\usepackage{jcappub}
 at 5truept 
\font\tmrms=ptmr at 6truept 
\input prepictex.tex
\input pictex.tex
\input postpictex.tex

\title{A Cosmological Slavnov-Taylor Identity}
\author{Hael Collins, }
\author{R.~Holman, }
\author{Tereza Vardanyan}
\affiliation{Physics Department, Carnegie Mellon University, Pittsburgh PA\ \ 15213, United States}
\emailAdd{hcollins@andrew.cmu.edu}
\emailAdd{rh4a@andrew.cmu.edu}
\emailAdd{tvardany@andrew.cmu.edu}

\abstract{
We develop a method for treating the consistency relations of inflation that includes the full time-evolution of the state.  This approach relies only on the symmetries of the inflationary setting, in particular a residual conformal symmetry in the spatial part of the metric, along with general properties which hold for any quantum field theory.  As a result, the consistency relations that emerge, which are essentially the Slavnov-Taylor identities associated with this residual conformal symmetry, apply very generally:  they are true of the {\it full\/} Green's functions, hold largely independently of the particular inflationary model, and can be used for {\it arbitrary\/} states.  We illustrate these techniques by showing the form assumed by the standard consistency relation between the two and three-point functions for the primordial scalar fluctuations when they are in a Bunch-Davies state.  But because we have included the full evolution of the state, this approach works for a general initial state as well and does not need to have assumed that inflation began in the Bunch-Davies state.  We explain how the Slavnov-Taylor identity is modified for these more general states.}

\arxivnumber{1405.0017}

\begin{document}
\maketitle

\setcounter{page}{2}

\section{Introduction} 

The theory of inflation provides an elegant causal explanation for the origin of the primordial fluctuations of the universe.  So far, the scalar component of these fluctuations has been the most carefully scrutinised, and its observation has richly added to our understanding of the early universe.  Planck and future experiments will continue to observe the more detailed structures in these scalar fluctuations, searching for further consistency between inflation and the observed universe.  And there are many other experiments underway working to measure the tensor component of the primordial fluctuations as well.  Because we are unlikely to be able to test the dynamics of inflation directly, finding such features in the primordial fluctuations and relations amongst them comes as an immense boon.  

One important prediction of inflation is that the Gaussian and non-Gaussian features in these primordial fluctuations should be related to each other.  In his analysis of the non-Gaussianities produced by inflation \cite{Maldacena:2002vr}, Maldacena presented a simple relation between the correlation functions of these fluctuations, $\zeta_{\vec k}(t)$.  His original relation states that in the limit where the momentum of one of the fields approaches zero, the three-point function of the fluctuations---the simplest measure of non-Gaussianity---is determined by the amplitude and scaling behaviour of the Gaussian power spectrum $P_k(t)$ through the identity
$$
\delta^3(\vec k_2+\vec k_3)\, P_{k\to 0}^{-1}(t_*)\, 
\langle\zeta_{\vec k\to\vec 0}(t_*)\zeta_{\vec k_2}(t_*)\zeta_{\vec k_3}(t_*)\rangle = \Bigl[ 3+{d\over d\ln k_2} \Bigr]\, P_{k_2}(t_*) . 
$$
Both sides of this expression are evaluated at a late time $t_*$ when the wavelengths responsible for the correlations seen in the cosmic microwave background and large-scale structure have been inflated to a size larger than the inflationary horizon.

Within a wide class of inflationary models---those with a single inflaton field whose potential obeys a set of `slow-roll' conditions, and where the fluctuations are in the Bunch-Davies state---this relation says that the amplitude of the three-point function in this `soft' limit should be small, since the basic properties of the power spectrum are already known through the observations made by Planck and its predecessors.  The smallness of the three-point function then follows naturally from assumptions that have already been made in these models, and the relation explains why the non-Gaussianities so far have been difficult to detect.  And once the three-point function has been measured, this relation will provide an important constraint on the consistency of the minimal inflationary picture.

Building on Maldacena's original insights, others have worked steadily during the past decade to generalise this relation, to extend it and to treat it from deeper perspectives \cite{Creminelli:2004yq,Creminelli:2012ed,Senatore:2012wy,Hinterbichler:2012nm,Hinterbichler:2013dpa,Berezhiani:2013ewa,Berezhiani:2014tda,Goldberger:2013rsa,Assassi:2012zq,Kehagias:2012pd,Schalm:2012pi,Pimentel:2013gza}.  It was shown, for example, that this consistency relation applies even when the slow-roll conditions are relaxed \cite{Creminelli:2004yq}, and other work \cite{Creminelli:2012ed,Senatore:2012wy} demonstrated that the original consistency relation is just the first in a series of relations between the $n+1$-point and $n$-point correlation functions predicted by inflation.

The fluctuations about the inflating background are quantum fields.  When the tensors are neglected, the scalar fluctuations can be cast, through a suitable choice of coordinates, in a form where they appear within a conformal factor that multiplies an otherwise flat spatial part of the metric.  This choice does not quite exhaust the diffeomorphism invariance of the metric.  Within this class of coordinates, conformal transformations of the spatial metric remain a residual symmetry of the theory.  What could be more natural then, than that these consistency conditions should be the Slavnov-Taylor identities associated with this symmetry of the quantum theory?

This realization was first developed by \cite{Hinterbichler:2012nm} and then refined and extended in later work \cite{Hinterbichler:2013dpa,Berezhiani:2013ewa,Berezhiani:2014tda}.  Recently Goldberger, Hui, and Nicolis \cite{Goldberger:2013rsa} have followed this approach to derive these relations using one-particle irreducible Green's functions and the effective action appropriate to an inflationary setting.  Since their approach relies only on very general properties of quantum field theory, it has great generality and the consistency relations follow as very simple and direct consequences of the residual conformal invariance.  Their derivation relies on only a minimal set of assumptions about the inflationary theory---that the initial state, its evolution, and the measure of the path integral are invariant under these conformal symmetries---which apply independently of slow-roll assumptions, the behaviour of the field outside the horizon, {\it etc\/}.  A further elegant feature of their derivation is that the path integral can be treated as that of a {\it three-dimensional\/} field theory defined on the late-time hypersurface, $t=t_*$.  Information about the initial state and its evolution influences a probability measure within the path integral; but as long as that measure remains invariant under the residual conformal symmetries, its details are not important for the Slavnov-Taylor identity for the fluctuations at the late-time boundary, $\zeta_{\vec k}(t_*)$.

An important element that has so far been missing from all previous treatments of the consistency relation is what happens when the scalar---or the tensor---fluctuations are {\it not\/} in a Bunch-Davies state.  One reason for considering states other than the Bunch-Davies state is that while inflation as an overall idea has been very successful in predicting many of the properties of the primordial fluctuations, we still know almost nothing about its details.  Cosmological observations are fine enough to be able now to exclude specific models, but it would be prudent to assume as little as we can and let observations constrain and tell us what is consistent with them.  Inflationary models are most often studied in terms of their potentials, but that is only half of the picture.  Allowing a more interesting initial state could help in explaining some of the observed properties of the primordial fluctuations.  Since it is straightforward to introduce more general initial states into a quantum field theory \cite{Agarwal:2012mq,Collins:2013kqa}, it should be possible to parametrise---and constrain by matching with what is seen---by how much the initial state of inflation can depart from the Bunch-Davies state in much the same way as we now study different actions for inflationary models in an effective theory language.

It is also essential to understand what any departures from the consistency relation could mean if they are seen.  If the three-point function is found not to be in accord with the na\"\i ve form of the consistency relation, it might be tempting to see this as a failure of the simple single-field, slow-roll picture for inflation.  But that is not the only possibility.  It could also be that the basic dynamical picture is correct, only that the state contains structures beyond the Bunch-Davies state; these too would modify the consistency relation.

What happens when the initial state breaks some of the residual conformal symmetry?  In this article we develop a method for calculating the cosmological Slavnov-Taylor identities for such states.  We shall still be following the philosophy of \cite{Goldberger:2013rsa}, except that now we allow the path integral to include the evolution of the state.  Our approach thus keeps the generality of the earlier approaches---which only relied on the symmetries of the action and the functional measure of the path integral---but it broadens them yet further by allowing for states that are not invariant under the full conformal symmetry.  Of course, the form of the consistency relation changes for a general state.  In fact these changes, though complicated in their expression, have a very familiar meaning:  they are the cosmological analogues for how explicitly broken symmetries alter the Ward identities in gauge theories.

The expression for the consistency relation that emerges from our approach can be used to treat much more general situations than those that have been considered so far.  To make some of the working details of our formalism clearer, we illustrate how the standard consistency relation, in the Bunch-Davies state, emerges before we undertake an analysis of a completely general initial state.  The corrections introduced into the Slavnov-Taylor identities by states that explicitly break the residual conformal symmetry are derived here; their phenomenology will be explored more fully in later work \cite{generalstate}. 

In the next section we introduce the action for an inflationary theory with a single scalar field.  What is most essential for the consistency relation are the symmetries that exist after we have chosen a particular set of coordinates or `gauge'.  When the coordinates are chosen so that the spatial part of the metric remains conformally flat, it is still left invariant by any further conformal transformation of the field.

In section 3 we develop the families of connected and one-particle irreducible Green's functions and the effective action that are needed for a fully time-evolving quantum field theory.  Once these theoretical tools have been introduced, we derive the Slavnov-Taylor identity and show that for spatial dilations it generates the consistency relation between the two and three-point functions.  While different expressions of this relation might be formally equivalent, it is most easily computed when it is written with a two-point 1PI Green's function acting on a three-point connected Green's function.  The necessary contortions needed to put it into this form are sketched in section 4 and explained more fully in an appendix.

Section 5 derives the form of the Slavnov-Taylor identity for a general initial state.  These states can be broadly divided according to whether they are invariant under the same conformal symmetries as the inflationary dynamics or whether they break some or all of these symmetries.  For the former states, although the Green's functions change, the consistency relations that relate them do not.  For the most general initial states, additional terms occur in the Slavnov-Taylor identity due to the non-invariance of the state.

The complete evaluation of the standard consistency relation for the simple class of inflationary theories analysed by Maldacena contains a few subtleties.  The singularities and zeros that occur in the soft momentum limit must be treated with care, so a detailed calculation of this relation for this standard---but extremely important---case is presented in appendix A.

In the conclusions we outline the next stages of this work.
\vskip24truept

\section{A residual diffeomorphism invariance} 

We begin by briefly reviewing the classical background used in inflation.  The simplest models of inflation contain a single scalar field $\phi(t,\vec x)$ with a potential $V(\phi)$.  The combined dynamics for gravity and this field are then determined by the action
$$
S = \int d^4\vec x\, \sqrt{-g} \Bigl\{ 
{\textstyle{1\over 2}}M_{\rm pl}^2R + {\textstyle{1\over 2}} g^{\mu\nu}\partial_\mu\phi\partial_\nu\phi - V(\phi) \Bigr\} , 
$$
where $M_{\rm pl}$ is the Planck mass.  Here the background space-time is assumed to depend only on the time coordinate.  In this case the classical value of the field can be written as just $\phi(t)$ and the background metric can be put into the standard form\footnote{Here we are using Maldacena's \cite{Maldacena:2002vr} notation for the background metric and the fluctuations about it.}
$$
ds^2 = dt^2 - e^{2\rho(t)}\, \delta_{ij}\, dx^idx^j . 
$$
The quantum fluctuations about this background very naturally introduce some spatial dependence into the universe during inflation.  As the time component already has a special role in this background, it is convenient to write the metric in the form
$$
ds^2 = \bigl[ N^2 - h_{ij}\, N^iN^j\bigr]\, dt^2 - 2h_{ij}\, N^i\, dtdx^j 
- h_{ij}\, dx^idx^j 
$$
to treat the fluctuations.  The background is spatially flat, so the quantum fluctuations are usually parametrised by how they transform under its symmetries.  Those that transform as spatial scalars are the most immediately observable since they are needed to explain the primordial inhomogeneities in the early universe.  The tensor fluctuations in $h_{ij}$ correspond to primordial gravity waves, which have also been recently observed.  Here we focus just on the scalar fluctuations, leaving the tensors for later work.

We use our freedom to choose our coordinates so that they have two useful properties:  first, that the inflaton field has no fluctuations at all, and is given entirely by its background value, $\phi(t,\vec x)=\phi(t)$, and second, that the spatial part of the metric remains proportional to $\delta_{ij}$.  The second of these conditions eliminates one of the scalar fluctuations in $h_{ij}$; the remaining scalar field $\zeta(t,\vec x)$ corresponds to quantum fluctuations in the scale factor itself,\footnote{If we had wished to include tensor fluctuations $\gamma_{ij}(t,\vec x)$ as well, we would have replaced $\delta_{ij}$ with 
$$
\delta_{ij} \to \exp[\gamma_{ij}] = \delta_{ij} + \gamma_{ij} 
+ {\textstyle{1\over 2}} \gamma_i^{\ k}\gamma_{kj} + \cdots .
$$}
$$
h_{ij} = e^{2\rho(t)+2\zeta(t,\vec x)}\, \delta_{ij} .
$$
This is an especially convenient choice since $\zeta(t,\vec x)$ is the quantum analogue of the fluctuations in the classical spatial curvature after inflation.  In the standard inflationary models it approaches a constant well outside the horizon.  The two further scalar fields present in $N$ and $N^i$ are fixed by the equations of motion for these fields, which are nondynamical Lagrange multipliers.  Written in terms of the field $\zeta(t,\vec x)$, the constraint equations found by varying the action with respect to $N$ and $N^i$ require that
$$
N = 1 + {\dot\zeta\over\dot\rho} 
\qquad\hbox{and}\qquad
N^i = \delta^{ij} \partial_j \biggl\{ - {e^{-2\rho}\over\dot\rho} \zeta + {1\over 2} {\dot\phi^2\over\dot\rho^2} \partial^{-2}\dot\zeta \biggr\} , 
$$
$\partial^{-2}$ being the inverse spatial Laplacian operator.

Is there any freedom left for changing the spatial coordinates further while still remaining within the general form that we have chosen?  As one instance of this, we observe that---since our background is flat---it should be possible to absorb a general conformal transformation of the spatial coordinates by a suitable change in $\zeta(t,\vec x)$.

Let us study this idea more precisely by making a transformation of the spatial coordinates 
$$
x^i \to x^i + \xi^i(t,\vec x) .
$$
We shall consider this to be a small transformation in the sense that it is consistent to work to linear order in the transformed quantities.  This transformation causes the spatial part of the metric to change by the amount
$$
\delta h_{ij} = e^{2\rho+2\zeta} \Bigl\{ 2\delta_{ij}\xi^k\partial_k\zeta 
+ \delta_{kj}\partial_i\xi^k + \delta_{kj}\partial_i\xi^k \Bigr\} ,
$$
to first order in $\xi^i$.  Once again we neglected the other fluctuations such as the tensor fluctuations.  This change can be absorbed through a corresponding change in the field $\zeta(t,\vec x)$, 
$$
\zeta(t,\vec x) \to \tilde\zeta(t,\vec x) 
= \zeta(t,\vec x) + \delta\zeta(t,\vec x) ,
$$
of the form 
$$
2\delta\zeta\, \delta_{ij} = 2\delta_{ij}\xi^k\partial_k\zeta 
+ \delta_{kj}\partial_i\xi^k + \delta_{kj}\partial_i\xi^k . 
$$
The fact that the change of coordinates can be related to an overall change in the factor multiplying the flat spatial metric means that $\xi^i(t,\vec x)$ is generating a conformal transformation.  We can make this property still more explicit by taking the trace of this equation, which allows us to solve for $\delta\zeta$ directly, 
$$\textstyle
\delta\zeta = \xi^k\partial_k\zeta + {1\over 3}\partial_k\xi^k , 
$$
and which, when substituted into the previous equation, yields the conformal Killing equation for the flat metric,
$$\textstyle
\partial_i\xi_j + \partial_j\xi_i = {2\over 3}\delta_{ij}\, \partial_k\xi^k .
$$

What we have found then is that even within the class of coordinates that we have chosen, there is some additional symmetry available.  The fluctuations $\zeta(t,\vec x)$ that we are considering are quantum fields, so these symmetries will generate constraints on the field and relations amongst its correlation functions.  Here we have found that our metric is invariant under the set of conformal transformations of three-dimensional flat space.  These transformations correspond to spatial translations and rotations, as well as dilations and special conformal transformations.  

Both the spatial translations and rotations have already implicitly been included in the structure of the theory, since $\zeta(t,\vec x)$ is itself a scalar field under these transformations.  And while the correlation functions of $\zeta(t,\vec x)$ must also be invariant under these symmetries, they do not impose any relations between correlation functions of different orders.  This follows from the fact that for both translations and rotations, the corresponding transformation of the field is {\it linear\/} and {\it homogeneous\/} in $\zeta(t,\vec x)$, $\delta\zeta = \xi^k\partial_k\zeta$, since $\partial_k\xi^k=0$.

In contrast, the dilation and the special conformal transformations introduce {\it inhomogeneous\/} terms into the infinitesimal transformation of the field.  For example, under a dilation,
$$
x^i \to x^i + \lambda\, x^i 
\qquad\hbox{or}\qquad 
\xi^i = \lambda\, x^i .
$$
The effect of this transformation on $\zeta(t,\vec x)$ is then 
$$\textstyle
\delta\zeta = \lambda + \lambda\, \vec x\cdot\vec\nabla\zeta .
$$
It is the presence of the inhomogeneous term that leads to relations between the $n$ and $n+1$-point correlation functions of the field.

While we shall not be treating the effects of a special conformal transformation further here, it too introduces an inhomogeneous term in the transformation of the scalar fluctuations.  A general special conformal transformation of the spatial coordinates is parametrised by a vector $\vec b$, 
$$
x^i \to {x^i-b^ix_kx^k\over 1-2b_kx^k+(b_jb^j)(x_kx^k)} .
$$
Under an infinitesimal transformation then, we have
$$
\xi^i = 2b^k\, x_kx^i - x^kx_kb^i .
$$
Putting this into the general expression for the corresponding transformation of $\zeta(t,\vec x)$, we arrive at a more complicated expression,
$$\textstyle
\delta\zeta = {2\over 3} \vec x\cdot\vec\nabla(\vec x\cdot\vec b)
- {1\over 3} |\!|\vec x|\!|^2 \vec\nabla\cdot\vec b 
+ {4\over 3} \vec x\cdot\vec b
+ 2(\vec x\cdot\vec b) \vec x\cdot\vec\nabla\zeta 
- |\!|\vec x|\!|^2 \vec b\cdot\vec\nabla\zeta . 
$$
This symmetry also generates relations between different orders of correlation functions of $\zeta(t,\vec x)$, since we can see that the first pair of terms are inhomogeneous and the second pair are linear in $\zeta(t,\vec x)$.

A fuller analysis of these residual symmetries, including a treatment of what additional time-dependent information can be gleaned by using the adiabatic properties \cite{Weinberg:2003sw} of the fluctuations at larger scales, is found in \cite{Hinterbichler:2012nm,Hinterbichler:2013dpa}.

\section{Evolution} 

The scalar fluctuation $\zeta(t,\vec x)$ is a quantum field whose dynamics are determined by the action $S[\zeta]$, which is found by expanding our original inflationary action in powers of $\zeta(t,\vec x)$.  In cosmological settings, the quantities that we should like to compute are the expectation values of operators built from this field.  In principle such expectation values could be taken in an arbitrary state, but in this article we shall always have a particular state in mind, the Bunch-Davies state, though we shall set up everything in a way that generalises fairly readily to other states.  The Bunch-Davies state is the state that matches with the free Minkowski vacuum in an arbitrarily remote past.  It is usually taken to be the natural choice for the ground state for inflation.  When the field is not in a free theory, its evolution from this pristine initial state set in an infinite past can be quite complicated and almost always needs to be treated perturbatively.

Let us write the expectation value of an operator ${\cal O}$ in the Bunch-Davies state, which we denote by $|0(t_*)\rangle$, as 
$$
\langle 0(t_*)|{\cal O}|0(t_*)\rangle .
$$
The state has been implicitly evolved from its initial free-theory form at $t=-\infty$ to its form at $t=t_*$, but we have been deliberately ambiguous about the time-dependence of the operator ${\cal O}$.  One of the most important expectation values for inflation is the equal-time correlation function of $n$ fields,
$$
\langle 0(t_*)|\zeta(t_*,\vec x_1)\cdots \zeta(t_*,\vec x_n)|0(t_*)\rangle . 
$$
However, to treat the full evolution of the theory we shall need more general $n$-point Green's functions too, where each field has its own independent time.

Before we construct these Green's functions, we ought first to explain how to treat the time-evolution of an expectation value in a little more detail.  Later, we shall be using the symmetries described in the previous section to derive relations amongst the expectation values of different $n$-point functions.  As we mentioned in the introduction, these relations are the Slavnov-Taylor identities adapted to a cosmological setting.  The path integral formalism is especially well suited for this purpose.  The generating functional for an evolving expectation value has the form,\footnote{We shall sometimes use these footnotes to explain what changes when we go from the Bunch-Davies state to a more general state---in preparation for the next stages of this work.  As a first comment:  for a more general state, the action will not necessarily have the diagonal structure, $S[\zeta^+]-S[\zeta^-]$.  It can have cross-terms where $\zeta^+$'s and $\zeta^-$'s couple directly to each other at an initial time $t_0$.}
$$
Z[J^\pm] = \int {\cal D}\zeta^+\,{\cal D}\zeta^-\, \exp\Bigl\{ 
iS[\zeta^+]-iS[\zeta^-] + i \int_{-\infty}^{t_*} dt\, \int d^3\vec x\, 
\bigl[ J^+\zeta^+ - J^-\zeta^- \bigr] \Bigr\}. 
$$
The overall structure of $Z[J^\pm]$ should be reminiscent of the generating function used in $S$-matrix calculations, except that here the fluctuation $\zeta(t,\vec x)$ has been written in terms of two fields, $\zeta^+(t,\vec x)$ and $\zeta^-(t,\vec x)$.  These two fields are associated with the evolution of the two states that occur within the expectation value.  The `$+$' index has been added to the field to signal that it is the time coordinate that occurs in the evolution of the state $|0(t)\rangle$ from its initial value, usually defined in some very remote past\footnote{Another---rather obvious---difference:  a general state could have been defined at an arbitrary initial time, $t_0$.}, up to the time $t_*$.  Similarly the $\zeta^-$ appears due to the evolution of the state $\langle 0(t)| = (|0(t)\rangle)^\dagger$.  This Hermitian conjugation introduces a few further oddities.  When time-ordering a product of fields, a $\zeta^-$ field always occurs {\it later\/} than any $\zeta^+$ field, whatever the numerical values of the times at which they occur might be.  This convention places time-ordered $\zeta^-$'s to the left, which is where they should be since they are associated with the $\langle 0(t)|$, which appears left-most in the expectation value.  The conjugation also reverses the evolution of time, so that the time-ordering of two $\zeta^-$ fields is the opposite of that of a pair of two $\zeta^+$ fields.  We shall write the explicit rules for the propagator below, where the time-ordering should become clearer.

As an example, let us show how the generating functional is used to define an equal-time $n$-point correlation function, 
$$
\langle 0(t_*)|\zeta(t_*,\vec x_1)\cdots \zeta(t_*,\vec x_n)|0(t_*)\rangle .
$$
Ordinarily, we generate such expectation values by applying $n$ functional derivatives with respect to the source to $Z[J^\pm]$.  But here we have two choices:  which sorts of fields---$\zeta^+$'s or $\zeta^-$'s---should these be?  The $\zeta$'s that appear here are meant to be `external fields.'  In that case it does not matter which sign we choose as long as they are {\it all\/} chosen to have the {\it same\/} sign.  What is conventionally done is to choose them to be all `$+$' fields.  The expectation value of $n$ fields at equal times is then evaluated---usually perturbatively---by computing the $n^{\rm th}$ functional derivative of $Z[J^\pm]$,
\begin{eqnarray}
&&\!\!\!\!\!\!\!\!\!\!\!\!\!\!\!\!\!
\langle 0(t_*)|\zeta(t_*,\vec x_1)\cdots \zeta(t_*,\vec x_n)|0(t_*)\rangle 
\nonumber \\
&&= 
\int {\cal D}\zeta^+\,{\cal D}\zeta^-\, 
\Bigl( \zeta^+(t_*,\vec x_1)\cdots \zeta^+(t_*,\vec x_n)\Bigr)
e^{ iS[\zeta^+]-iS[\zeta^-] + i \int_{-\infty}^{t_*} dt\, \int d^3\vec x\, 
[J^+\zeta^+ - J^-\zeta^-] }. 
\nonumber \\
&&= 
\biggl( {1\over i}{\delta\over\delta J^+(t_*,\vec x_1)} \biggr)\cdots 
\biggl( {1\over i}{\delta\over\delta J^+(t_*,\vec x_n)} \biggr) 
Z[J^\pm]\Bigr|_{J^\pm=0} . 
\nonumber 
\end{eqnarray}
These equal-time correlation functions are what inflation is meant to generate.  They are the initial conditions that the inflationary era bequeaths to the subsequent eras.

Once we have introduced the generating functional, we can define further Green's functions which are needed in deriving the cosmological Slavnov-Taylor identities.  Consider a family of Green's functions evaluated now at arbitrary space-time points and with arbitrary $\pm$ indices too, 
$$
G^{\pm_1\cdots\pm_n}(x_1,\ldots,x_n) 
= \langle 0(t_*)| T\bigl( \zeta^{\pm_1}(x_1) \cdots \zeta^{\pm_n}(x_n) \bigr)
|0(t_*)\rangle .
$$
They are defined by taking functional derivatives of the path integral with respect to the appropriate sources, $J^\pm(x_n)$, 
$$
G^{\pm_1\cdots\pm_n}(x_1,\ldots,x_n) 
= \biggl( {1\over \pm i}{\delta\over\delta J^{\pm_1}(x_1)} \biggr)
\cdots 
\biggl( {1\over \pm i}{\delta\over\delta J^{\pm_n}(x_n)} \biggr) 
Z[J^\pm]\Bigr|_{J^\pm=0} . 
$$
In principle, in addition to the time-dependence of each field, $x_i=(t_i,\vec x_i)$, there is a further time-dependence\footnote{or rather, a dependence on {\it two\/} times, which we could call $t_*$ and $t_*'$,
$$
\langle 0(t_*')|{\cal O}|0(t_*)\rangle 
= \int {\cal D}\zeta^+\,{\cal D}\zeta^-\, {\cal O} \exp\Bigl\{ 
i\int_{-\infty}^{t_*} dt\, \int d^3\vec x\, \bigl[ {\cal L}[\zeta^+] + J^+\zeta^+\bigr] 
- i\int_{-\infty}^{t_*'} dt\, \int d^3\vec x\, \bigl[ {\cal L}[\zeta^-] + J^-\zeta^-\bigr] \Bigr\} ,
$$
but we shall refrain from treating anything so perverse here.} given by the time to which the states are being evolved, which we have called $t_*$,
$$
G^{\pm_1\cdots\pm_n}(x_1,\ldots,x_n)=G^{\pm_1\cdots\pm_n}(t_*;x_1,\ldots,x_n) . 
$$
Since the Green's functions here will always be evaluated in a state at the $t=t_*$ hypersurface, this dependence will not be written explicitly.  We shall always assume that $t_i\le t_*$.

Green's functions that contain a mixture of $+$ and $-$ indices and that depend on multiple times do not correspond to graphs where the fields are all `external'---the analogue of `on-shell' external fields in an $S$-matrix calculation.  However, such Green's functions naturally occur as the internal subgraphs of a more complicated process.  When we integrate the position of an internal vertex over all space-time points, this also means that here we are summing over $\pm$ indices.\footnote{A note for the experts:  if we think instead of the $t$ in the $\zeta^+(t,\vec x)$ and $\zeta^-(t,\vec x)$ as being a {\it single\/} time coordinate $t_c$ that runs along a contour from the initial to the final time and then back again, the sum over $\pm$ indices naturally appears when we integrate internal vertices of the fields $\zeta(t_c,\vec x)$ over the {\it entire\/} time contour in $dt_cd\vec x$.}

The simplest examples of such internal Green's functions are the Feynman propagators themselves.  They are derived from the quadratic part of the action and they form the basis of a perturbative treatment of more complicated processes.  In taking the Wick contractions of pairs of fields, there are four possibilities for the $\pm$ indices of the two fields that are contracted, which in turn means that there are four types of Feynman propagators,
$$
G^{\pm\pm}(x,y) 
= \int {d^3\vec k\over (2\pi)^3}\, e^{i\vec k\cdot(\vec x-\vec y)}\, 
G_k^{\pm\pm}(t,t') ,
$$
with $x=(t,\vec x)$ and $y=(t',\vec y)$ and where
\begin{eqnarray}
G_k^{++}(t,t') &=& 
\Theta(t-t')\, G_k^>(t,t') + \Theta(t'-t)\, G_k^<(t,t') 
\nonumber \\
G^{+-}(t,t') &=& G_k^<(t,t') 
\nonumber \\
G^{-+}(t,t') &=& G_k^>(t,t') 
\nonumber \\
G^{--}(t,t') &=& 
\Theta(t'-t)\, G_k^>(t,t') + \Theta(t-t')\, G_k^<(t,t') . 
\nonumber 
\end{eqnarray}
Here, $G^>_k(t,t')$ and $G^<_k(t,t')$ are the free Wightman functions associated with the two-point functions evaluated in the asymptotic vacuum state,\footnote{For an initial state defined at $t_0$, they would be evaluated in that state, $|0(t_0)\rangle$, instead.} 
\begin{eqnarray}
\langle 0(-\infty)|\zeta(t,\vec x)\zeta(t',\vec y)|0(-\infty)\rangle 
&=& 
\int {d^3\vec k\over (2\pi)^3}\, e^{i\vec k\cdot(\vec x-\vec y)}\, G_k^>(t,t') 
\nonumber \\
\langle 0(-\infty)|\zeta(t',\vec y)\zeta(t,\vec x))|0(-\infty)\rangle 
&=& 
\int {d^3\vec k\over (2\pi)^3}\, e^{i\vec k\cdot(\vec x-\vec y)}\, G_k^<(t,t') .
\nonumber 
\end{eqnarray}

Notice that the time-ordering follows the pattern that we described before.  Times associated with a $-$ field always occur after those of a $+$ field, which explains the absence of $\Theta$-functions in $G_k^{+-}(t,t')$ and $G_k^{-+}(t,t')$, and the fact that the time-ordering of the $-$ fields is the opposite of that of the $+$ fields is a relic of the Hermitian conjugation.  This also explains the reversed roles of the $\Theta$-functions in $G_k^{--}(t,t')$.

What are these Wightman functions for an inflationary universe?  The free, or quadratic, part of the action for a slow-roll model of inflation is 
$$
S^{(2)}[\zeta] = {1\over 2} \int_{-\infty}^{t_*} dt\, e^{3\rho(t)} {\dot\phi^2\over\dot\rho^2} \int d^3\vec x\, 
\Bigl\{ \dot\zeta^2 - e^{-2\rho(t)}\partial_k\zeta\partial^k\zeta \Bigr\} . 
$$
The slow-roll limit of inflation corresponds to the regime where the dimensionless slow-roll parameters 
$$
\epsilon = {1\over 2}{1\over M_{\rm pl}^2} {\dot\phi^2\over\dot\rho^2}
\qquad\hbox{and}\qquad
\delta = {\ddot\phi\over\dot\rho\dot\phi} 
$$
are small:  $\epsilon,\delta\ll 1$.  In this limit, the Wightman functions derived from these quadratic terms are
$$
G^>_k(t,t') = G^<_k(t',t') = e^{-\rho(t)} e^{-\rho(t')} {\dot\rho(t)\over\dot\phi(t)} {\dot\rho(t')\over\dot\phi(t')} 
{\pi\over 4} \sqrt{\eta\eta'} H_\nu^{(1)}(-k\eta)H_\nu^{(2)}(-k\eta') ,
$$
where the index of the Hankel functions is
$$\textstyle
\nu= \sqrt{{9\over 4}+3(2\epsilon+\delta)} . 
$$
Here some of the time-dependence has been written in terms of the conformal time, $\eta=\eta(t)$ and $\eta'=\eta(t')$, where
$$
\eta(t) = \int dt\, e^{-\rho(t)} , 
$$
since the expressions then look simpler.  In an inflationary universe the conformal time is usually chosen to be negative, $\eta\in(-\infty,0)$, since for this choice the coordinate runs forward when the time is also running forward.  The leading terms of these Wightman functions in the slow-roll limit have a simpler, approximately de Sitter, form
\begin{eqnarray}
G^>_k(t,t') &=& 
{1\over 4\epsilon} {H^2\over M_{\rm pl}^2} {1\over k^3} 
(1+ik\eta)(1-ik\eta') e^{-ik(\eta-\eta')} + \cdots
\nonumber \\
G^<_k(t,t') &=& 
{1\over 4\epsilon} {H^2\over M_{\rm pl}^2} {1\over k^3} 
(1-ik\eta)(1+ik\eta') e^{ik(\eta-\eta')} + \cdots . 
\nonumber 
\end{eqnarray}

Just as we can perturbatively treat a general diagram as a graph of $G^{\pm\pm}(x,y)$ propagators connecting vertices composed of just $+$ or $-$ fields, we can similarly imagine more complicated diagrams as being divided into various subgraphs.  When a line of a subgraph does not end at an external space-time point, it can be at an arbitrary time with an arbitrary $\pm$ index, though when summing up the graphs that contribute to a process we integrate over the space-time location of the internal point {\it and\/} sum over all values of $\pm$.  We shall see later that the consistency relation has this structure.

The Green's functions that we have introduced so far correspond to the sums of {\it all\/} the graphs that contribute to a process, whether they are connected or not.\footnote{We did not need to remove the vacuum-to-vacuum graphs explicitly when we defined these Green's functions.  The vacuum-to-vacuum graphs are automatically cancelled due to the $S[\zeta^+]-S[\zeta^-]$ structure that appears in $Z[J^\pm]$.}  For the two and three-point functions, we do not need to distinguish between the connected and unconnected Green's functions in this theory.  However, for higher point functions, the Green's functions can be separated into a sum of products of lower-order $n$-point functions plus purely connected components.  For example, the four-point function can be written as a sum of three pairs of two-point functions plus a connected four-point function, {\it e.g.\/}
\begin{eqnarray}
G^{++++}(x_1,x_2,x_3,x_4) &=& 
G^{++}(x_1,x_2) G^{++}(x_3,x_4) 
+ G^{++}(x_1,x_3) G^{++}(x_2,x_4) 
\nonumber \\
&&
+\,\, G^{++}(x_1,x_4) G^{++}(x_2,x_3) 
+ G_c^{++++}(x_1,x_2,x_3,x_4) .
\nonumber 
\end{eqnarray}
We can extract these latter from the rest by defining a generating functional, $W[J^\pm]$, for just the {\it connected\/} graphs, 
$$
Z[J^\pm] = e^{iW[J^\pm]} .
$$
The connected $n$-point Green's functions are then defined by taking the appropriate functional derivatives with respect to $W[J^\pm]$,
$$
G^{\pm_1\cdots\pm_n}_c(x_1,\ldots,x_n) 
= \biggl( {1\over \pm i} {\delta\over\delta J^{\pm_1}(x_1)} \biggr)\cdots 
\biggl( {1\over \pm i} {\delta\over\delta J^{\pm_n}(x_n)} \biggr) 
iW[J^\pm]\Bigr|_{J^\pm=0} . 
$$
Graphically, 
$$
\beginpicture
\setcoordinatesystem units <1.00truept,1.00truept>
\setplotarea x from -72 to 48, y from -16 to 16
\circulararc 360 degrees from 16 0 center at 0 0
\plot  45.11  16.42   15.04  5.47 /
\plot  47.27   8.34   15.76  2.78 /
\plot  45.11 -16.42   15.04 -5.47 /
\plot -45.11  16.42  -15.04  5.47 /
\plot -47.27   8.34  -15.76  2.78 /
\plot -45.11 -16.42  -15.04 -5.47 /
\put {$G^{\pm_1\cdots\pm_n}_c(x_1,\ldots,x_n) = $} [r] at -80 0
\put {$\cdot$} [c] at 32.00  0.00
\put {$\cdot$} [c] at 31.88 -2.79
\put {$\cdot$} [c] at 31.51 -5.56
\put {$\cdot$} [c] at -32.00  0.00
\put {$\cdot$} [c] at -31.88 -2.79
\put {$\cdot$} [c] at -31.51 -5.56
\put {{\scriptsize $x_1,\pm_1$}} [r] at -46  16.42
\put {{\scriptsize $x_2,\pm_2$}} [r] at -49   8.34
\put {{\scriptsize $x_i,\pm_i$}} [r] at -46 -16.42
\put {{\scriptsize $x_n,\pm_n$}} [l] at  48  16.42
\put {{\scriptsize $x_{n-1},\pm_{n-1}$}} [l] at 51  8.34
\put {{\scriptsize $x_{i+1},\pm_{i+1}$}} [l] at 48 -16.42
\setshadesymbol ({\tmrms .})
\setshadegrid span <0.9pt>
\setquadratic
\hshade -11.3 -11.3 -11.3 <,z,,>  0 -16.0 -11.3   11.3 -11.3 -11.3 /
\vshade -11.3 -11.3  11.3 <z,z,,> 0 -16.0  16.0   11.3 -11.3  11.3 /
\hshade -11.3  11.3  11.3 <z,,,>  0  11.3  16.0   11.3  11.3  11.3 /
\endpicture
$$
where the shaded blob is the sum of all connected diagrams.

Proceeding a step further, we introduce a generating functional for the `one-particle irreducible' Green's function through a Legendre transform of the $W[J^\pm]$.  Let us define the connected expectation value of the field $\zeta(t,\vec x)$ in the presence of a source to be 
$$
\bar\zeta^\pm(x) \equiv \pm {\delta W\over\delta J^\pm(x)} 
= \langle 0(t)|\zeta^\pm(x)|0(t)\rangle_{c,J^\pm} . 
$$
The functional $\Gamma[\bar\zeta^\pm]$ is then defined through the transformation, 
$$
\Gamma[\bar\zeta^\pm] = W[J^\pm] - \int d^4x\, 
\bigl[ J^+(x)\bar\zeta^+(x) - J^-(x)\bar\zeta^-(x) \bigr] . 
$$
If we take the functional derivative of $\Gamma[\bar\zeta^\pm]$ with respect to $\bar\zeta^\pm$, we produce the complementary relations 
$$
J^\pm(x) = \mp {\delta\Gamma\over\delta\bar\zeta^\pm(x)} . 
$$
The individual `1PI' $n$-point functions are defined by differentiating with respect to the fields $\bar\zeta^\pm$ an appropriate number of times,
$$
\Gamma^{\pm_1\cdots\pm_n}(x_1,\ldots,x_n) 
= {\delta^n \Gamma[\bar\zeta^\pm]\over\delta\bar\zeta^{\pm_1}(x_1) \cdots\delta\bar\zeta^{\pm_n}(x_n)} \biggr|_{\bar\zeta^\pm=0} . 
$$
We might have defined this equation with some convention for the signs too, but it is simpler in this instance not to do so.

The 1PI functional $\Gamma[\bar\zeta^\pm]$ is also called the {\it effective action\/}.  For a renormalizable theory, it is assumed that the effective vertices, $\Gamma^{\pm_1\cdots\pm_n}(x_1,\ldots,x_n)$, have all been renormalized, when it is necessary to do so.  The connected diagrams associated with the Green's functions $G_c^{\pm_1\cdots\pm_n}(x_1,\ldots,x_n)$, and which are calculated from all the tree and loop diagrams generated by the original action $S[\zeta^\pm]$, correspond to the tree-diagrams calculated using the effective action $\Gamma[\bar\zeta^\pm]$.

In the Bunch-Davies state, many of the 1PI effective vertices vanish.  The structure of the effective action $\Gamma[\bar\zeta^\pm]$ mirrors, in part, the structure of the original action
$$
S[\zeta^+] - S[\zeta^-] , 
$$
which is composed entirely of operators containing just the $\zeta^+(t,\vec x)$ or just the $\zeta^-(t,\vec x)$ field, but with no operators coupling the two.  Certainly any effective vertex that requires a counterterm of the same order will need also to have this same structure---for otherwise there would not have been the possibility of such a counterterm in the original action.  This suggests that the effective vertices should vanish except when all of the indices are $+$ or all are $-$, and that they are related by a single sign, 
$$
\Gamma^{(n)}(x_1,\ldots,x_n) \equiv \Gamma^{+\cdots +}(x_1,\ldots,x_n) 
= - \Gamma^{-\cdots -}(x_1,\ldots,x_n) .
$$
Then the effective action becomes
$$
\Gamma[\bar\zeta^\pm] = \Gamma_{\hbox{\tiny BD}}[\bar\zeta^+] 
- \Gamma_{\hbox{\tiny BD}}[\bar\zeta^-] ,
$$
where
$$
\Gamma_{\hbox{\tiny BD}}[\bar\zeta] 
= \sum_{n=2}^\infty {1\over n!} \int d^4x_1\cdots d^4x_n\, 
\Gamma^{(n)}(x_1,\ldots, x_n)\, \bar\zeta(x_1) \cdots \bar\zeta(x_n) . 
$$

This is {\it not true\/} of a more general ground state.  An initial state can be defined through an initial action defined on a $t=t_0$ hypersurface \cite{Agarwal:2012mq,Collins:2013kqa}.  A general initial action {\it does\/} contain operators that couple the $\zeta^+$ and $\zeta^-$ fields directly, so the action no longer has the simple $S[\zeta^+]-S[\zeta^-]$ form of a Bunch-Davies state.  Correspondingly, the effective action will also contain effective vertices with mixed $\pm$ indices.

Apart from the addition of the $\pm$ indices, all of these structures have their familiar analogues in $S$-matrix calculations.  There is, however, a further difference here which makes calculations in this setting slightly more cumbersome.  In a typical scattering problem, we are evolving a system from $t=-\infty$ to $\infty$.  When we perform a Fourier transform, it is then in all $3+1$ dimensions, and the resulting Green's functions typically have simpler structures when expressed in terms of the four-momenta, rather than in terms of the original space-time coordinates.  In a cosmological setting we are evolving only up to a finite time $t_*$, and we could have started from a finite initial time $t_0$ too.  Therefore we shall only be Fourier transforming in the spatial dimensions, $(t,\vec x)\to(t,\vec k)$.  For example, the two-point 1PI Green's function remains a differential operator rather than becoming a purely algebraic function.

Let us illustrate this last point more fully with a particular example, which we shall need later anyway.  The 1PI and the connected two-point functions are the functional inverses of each other.  This relation follows from 
$$
\int d^4z\, \biggl\{ {\delta J^+(y)\over\delta\bar\zeta^+(z)}
{\delta\bar\zeta^+(z)\over\delta J^+(x)}
+ {\delta J^+(y)\over\delta\bar\zeta^-(z)}
{\delta\bar\zeta^-(z)\over\delta J^+(x)}
\biggr\}
= \sum_{s=\pm} \int d^4z\, {\delta J^+(y)\over\delta\bar\zeta^s(z)}
{\delta\bar\zeta^s(z)\over\delta J^+(x)}
= \delta^4(x-y) ,
$$
for example.  Here we have shown the case when both the sources are $J^+$'s; but we could have just as well written a similar relation for any choice of the signs,
$$
\sum_{s=\pm} \int d^4z\, {\delta J^{r_2}(y)\over\delta\bar\zeta^s(z)}
{\delta\bar\zeta^s(z)\over\delta J^{r_1}(x)}
= \delta_{r_1}^{r_2}\, \delta^4(x-y) , 
$$
where $r_1$ and $r_2$ can each assume either sign.  This relation could also have  been expressed as the functional derivatives of the appropriate generating functional, 
$$
-r_2\sum_{s=\pm} s\int d^4z\, {\delta^2W\over\delta J^{r_1}(x)\delta J^s(z)}
{\delta^2\Gamma\over\delta\bar\zeta^s(z)\delta\bar\zeta^{r_2}(y)}
= \delta_{r_1}^{r_2}\, \delta^4(x-y) ,
$$
which in turn becomes a relation between the two-point functions, 
$$
\sum_{s=\pm} \int d^4z\, G_c^{r_1s}(x,z)\Gamma^{sr_2}(z,y)
= i\delta_{r_1}^{r_2}\, \delta^4(x-y) .
$$
Fourier transforming in the spatial dimensions, 
\begin{eqnarray}
G_c^{rs}(t,\vec x;t',\vec y) &=& 
\int {d^3\vec k\over (2\pi)^3}\, e^{i\vec k\cdot(\vec x-\vec y)} G_k^{rs}(t,t') 
\nonumber \\
\Gamma^{rs}(t,\vec x;t',\vec y) &=& 
\int {d^3\vec k\over (2\pi)^3}\, 
e^{i\vec k\cdot(\vec x-\vec y)} \Gamma_k^{rs}(t,t') ; 
\nonumber 
\end{eqnarray}
the statement that the two types of two-points functions are each other's functional inverse corresponds in this setting to the four equations
\begin{eqnarray}
\int_{-\infty}^{t_*} dt^{\prime\prime}\, \Bigl\{
\Gamma^{++}_k(t,t^{\prime\prime})\, G^{++}_k(t^{\prime\prime},t') 
+ \Gamma^{+-}_k(t,t^{\prime\prime})\, G^{-+}_k(t^{\prime\prime},t')
\Bigr\} &=& i\, \delta(t-t') 
\nonumber \\
\int_{-\infty}^{t_*} dt^{\prime\prime}\, \Bigl\{
\Gamma^{++}_k(t,t^{\prime\prime})\, G^{+-}_k(t^{\prime\prime},t') 
+ \Gamma^{+-}_k(t,t^{\prime\prime})\, G^{--}_k(t^{\prime\prime},t')
\Bigr\} &=& 0 
\nonumber \\
\int_{-\infty}^{t_*} dt^{\prime\prime}\, \Bigl\{
\Gamma^{--}_k(t,t^{\prime\prime})\, G^{-+}_k(t^{\prime\prime},t')
+ \Gamma^{-+}_k(t,t^{\prime\prime})\, G^{++}_k(t^{\prime\prime},t') 
\Bigr\} &=& 0 
\nonumber \\
\int_{-\infty}^{t_*} dt^{\prime\prime}\, \Bigl\{
\Gamma^{--}_k(t,t^{\prime\prime})\, G^{--}_k(t^{\prime\prime},t')
+ \Gamma^{-+}_k(t,t^{\prime\prime})\, G^{+-}_k(t^{\prime\prime},t') 
\Bigr\} 
&=& i\, \delta(t-t') . 
\nonumber 
\end{eqnarray}

These equations remain true for an arbitrary initial state,\footnote{or rather, one that is not quite completely arbitrary---we are still assuming that the spatial translations and rotations are unbroken.  Importantly, however, we do not need to assume that the state is invariant under dilations or special conformal transformations.} with only the tiny modification that if we had fixed the state at $t_0$, the lower limits would have been replaced by $t_0$.  In the Bunch-Davies state, we can additionally assume that 
$$
\Gamma^{+-}_k(t,t^{\prime\prime}) = 0,
\qquad\hbox{and}\qquad
\Gamma^{-+}_k(t,t^{\prime\prime}) = 0.
$$
These functional relations are between the {\it full\/} Green's functions; that is, the sums of {\it all\/} connected or 1PI graphs with the appropriate external structures.  For the calculation of the consistency relation, we only require their leading behaviour.  In that limit, the connected two-point functions are then just the free Feynman propagators and the 1PI two-point functions are the operators derived from the quadratic equation of motion for the field $\zeta_k(t)$,
$$
\Gamma^{++}_k(t,t')= - \Gamma^{--}_k(t,t') = - \delta(t-t') \biggl\{ 
{\dot\phi^2\over\dot\rho^2} e^{3\rho} {d^2\over dt^{\prime 2}} 
+ {d\over dt'}\biggl[ {\dot\phi^2\over\dot\rho^2} e^{3\rho} \biggr] {d\over dt'}
+ k^2 {\dot\phi^2\over\dot\rho^2} e^\rho \biggr\} . 
$$

\section{A Slavnov-Taylor identity} 

Now, we are ready to derive a consistency condition between the two-point and three-point Green's functions by putting these ideas together.  As our starting point, we use the invariance of the functional measure of the generating functional $Z[J^\pm]$ under a transformation of the fields,
$$
\zeta(t,\vec x) \to \tilde\zeta(t,\vec x) 
= \zeta(t,\vec x) + \delta\zeta(t,\vec x) .
$$
Here, we are assuming that $\delta\zeta(t,\vec x)$ is linear in the field, with the possibility of an inhomogeneous term too.  When $\delta\zeta(t,\vec x)$ has this form, the measure of the functional integral is invariant ${\cal D}\tilde\zeta={\cal D}\zeta$.  So under this change of the functional integration variable, $Z[J^\pm]$ has not changed, 
\begin{eqnarray}
Z[J^\pm] &=& 
\int {\cal D}\zeta^+\,{\cal D}\zeta^-\, 
e^{ iS[\zeta^+]-iS[\zeta^-] + i \int_{-\infty}^{t_*} dt\, \int d^3\vec x\, 
[J^+\zeta^+ - J^-\zeta^-] } 
\nonumber \\
&=& 
\int {\cal D}\tilde\zeta^+\,{\cal D}\tilde\zeta^-\, 
e^{ iS[\tilde\zeta^+]-iS[\tilde\zeta^-] + i \int_{-\infty}^{t_*} dt\, \int d^3\vec x\, 
[J^+\tilde\zeta^+ - J^-\tilde\zeta^-] } . 
\nonumber 
\end{eqnarray}
When the transformation that we have made is additionally a symmetry of the action, as is the case for the conformal transformations mentioned earlier, we have that $S[\tilde\zeta] = S[\zeta]$, together with the invariance of the functional measure, yields
$$
Z[J^\pm] = \int {\cal D}\zeta^+\,{\cal D}\zeta^-\, 
e^{ iS[\zeta^+]-iS[\zeta^-] + i \int_{-\infty}^{t_*} dt\, \int d^3\vec x\, 
[J^+(\zeta^+ + \delta\zeta^+) - J^-(\zeta^- + \delta\zeta^-)] } . 
$$
The only remnant of the transformation appears in its coupling to the sources.  Under a small transformation, we can expand the exponential to linear order in $\delta\zeta^\pm$,
$$
Z[J^\pm] = Z[J^\pm] + \delta Z[J^\pm] ,
$$
where 
$$
\delta Z[J^\pm] = i \int_{-\infty}^{t_*} dt\, \int d^3\vec x\, 
\Bigl[ J^+(t,\vec x)\, \langle\delta\zeta^+(t,\vec x)\rangle_{J^\pm} 
- J^-(t,\vec x)\, \langle\delta\zeta^-(t,\vec x)\rangle_{J^\pm} \Bigr] . 
$$
The $\langle\delta\zeta^\pm(t,\vec x)\rangle_{J^\pm}$ are the expectation values of the infinitesimal change in the field in the presence of the source $J^\pm$.  Because the generating functional has not changed, we conclude that 
$$
\delta Z[J^\pm] = 0 .
$$
This result is called the {\it Slavnov-Taylor identity\/}.\footnote{The derivation of the Slavnov-Taylor identity that we have presented here, essentially follows the reasoning of \cite{Weinberg:1996kr}.}

The next step is to use this identity for the residual conformal symmetry to generate relations amongst Green's functions of different orders.  The most important of these relations for current observations is the one associated with the spatial dilations.  Under a dilation, we found that the field $\zeta(t,\vec x)$ changes infinitesimally by an amount 
$$
\delta\zeta = \lambda\bigl[ 1 + \vec x\cdot\vec\nabla\zeta \bigr] .
$$
For a dilation then, the Slavnov-Taylor identity becomes
$$
\int_{-\infty}^{t_*} dt\, \int d^3\vec x\, \biggl[ 
J^+(t,\vec x) - J^-(t,\vec x)
+ J^+(t,\vec x)\, \vec x\cdot\vec\nabla 
  {\delta W[J^\pm]\over\delta J^\pm(t,\vec x)}
+ J^-(t,\vec x)\, \vec x\cdot\vec\nabla 
  {\delta W[J^\pm]\over\delta J^\pm(t,\vec x)} \biggr] = 0 . 
$$
Here we have rewritten the one-point function in the presence of a source as
$$
\langle\zeta^\pm(t,\vec x)\rangle_{J^\pm} 
= {1\over\pm i}{\delta Z[J^\pm]\over\delta J^\pm(t,\vec x)}
= \pm Z[J^\pm] {\delta W[J^\pm]\over\delta J^\pm(t,\vec x)} ,
$$
and we have removed some non-vanishing prefactors.  By differentiating this relation $n$ times with respect to $\bar\zeta^\pm$ we generate relations between the $n+1$ and $n$-point functions.

While the higher-order relations are useful since they place constraints on the (so far) unobserved higher-order correlation functions of the primordial fluctuations, the most immediately important of these relations for observations is that between the three and two-point functions.  The two-point function is the only correlation function that has been observed so far, and a decisive measurement of the three-point function is being sought.  This relation then places a strong constraint on the amplitude of the three-point function if it is to be consistent with the simplest class of inflationary models with only one inflaton field and where the potential satisfies the slow-roll conditions.

The calculation of this standard consistency relation goes as follows:  (1) first differentiate the Slavnov-Taylor dilation identity twice with respect to $\bar\zeta^\pm$, setting $\bar\zeta^\pm=0$ after doing so, and then (2) move some of the two-point functions from one set of terms to the other.  The second step is more aesthetic---we essentially have the consistency relation after the first step---but it is useful nonetheless since it puts the relation in a form more closely resembling its usual expression elsewhere; and perhaps more importantly, it is only finite and nonzero on both sides of its equation after this second step.

So let us start by differentiating the Slavnov-Taylor identity with respect to the fields $\bar\zeta^{s_2}(y_2)$ and $\bar\zeta^{s_3}(y_3)$, where we have given them arbitrary $s_2,s_3=\pm$ indices, to produce 
\begin{eqnarray}
&&\!\!\!\!\!\!\!\!\!\!\!\!\!\!\!\!
\int d^4x\, \biggl\{ 
{\delta^2 J^+(x)\over\delta\bar\zeta^{s_2}(y_2)\delta\bar\zeta^{s_3}(y_3)} 
- {\delta^2 J^-(x)\over\delta\bar\zeta^{s_2}(y_2)\delta\bar\zeta^{s_3}(y_3)} 
\nonumber \\
&&\quad
+ \sum_{r,s=\pm} \int d^4z\, {\delta J^r(x)\over\delta\bar\zeta^{s_2}(y_2)}
{\delta J^s(z)\over\delta\bar\zeta^{s_3}(y_3)} \vec x\cdot\vec\nabla_{\vec x} {\delta W\over\delta J^r(x)\delta J^s(z)} 
\nonumber \\
&&\quad
+ \sum_{r,s=\pm} \int d^4z\, {\delta J^r(x)\over\delta\bar\zeta^{s_3}(y_3)}
{\delta J^s(z)\over\delta\bar\zeta^{s_2}(y_2)}
\vec x\cdot\vec\nabla_{\vec x} {\delta W\over\delta J^r(x)\delta J^s(z)} 
\biggr\} = 0 . 
\nonumber 
\end{eqnarray}
To keep the expressions more compact, we have again combined the time and space coordinates and written the arguments as a single four-vector, {\it e.g.\/}~$x=(t,\vec x)$.

The first pair of terms is not quite what we want.  Cosmological observations are used to constrain the {\it connected\/} three-point functions of the primordial fluctuations.  What we have here are instead the 1PI three-point functions.  If we recall how the functional derivatives of $J^\pm$ and $W$ are related to the 1PI and the connected Green's functions, the above equation can be expressed diagrammatically as 
\begin{eqnarray}
\beginpicture
\setcoordinatesystem units <1.00truept,1.00truept>
\setplotarea x from -72 to 48, y from -32 to 32
\circulararc 360 degrees from 12 0 center at 0 0
\plot -36 0  -12 0 /
\plot  6  10.39  18  31.18 /
\plot  6 -10.39  18 -31.18 /
\put {{\footnotesize $1PI$}} [c] at 0 0
\put {$\displaystyle -i\sum_{s=\pm}\int dtd^3\vec x\, $} [r] at -40 0
\put {{\scriptsize $t, \vec x, s$}} [c] at -36 -7
\put {{\scriptsize $t_2, y_2, s_2$}} [l] at 20  27
\put {{\scriptsize $t_3, y_3, s_3$}} [l] at 20 -30
\endpicture
&=&
\beginpicture
\setcoordinatesystem units <1.00truept,1.00truept>
\setplotarea x from -60 to 40, y from -12 to 12
\circulararc 360 degrees from  12 0 center at 0 0
\circulararc 360 degrees from -34 0 center at -42 0
\circulararc 360 degrees from  22 0 center at  30 0
\plot -50 0  -60 0 /
\plot -27 0  -34 0 /
\plot -21 0  -12 0 /
\plot  12 0   22 0 /
\plot  38 0   48 0 /
\put {{\tiny $1PI$}} [c] at -42 0
\put {{\tiny $1PI$}} [c] at  30 0
\put {{\scriptsize $\otimes$}} [c] at -24 0
\put {{\tiny $\vec x\cdot\vec\nabla_{\vec x}$}} [c] at -23 8
\put {{\tiny $t_2, y_2, s_2$}} [c] at -66 -7
\put {{\tiny $t_3, y_3, s_3$}} [c] at  54 -7
\setshadesymbol ({\tmrms .})
\setshadegrid span <0.9pt>
\setquadratic
\hshade -8.475 -8.475 -8.475 <,z,,>  0 -12.0   -8.475 8.475 -8.475 -8.475 /
\vshade -8.475 -8.475  8.475 <z,z,,> 0 -12.0   12.0   8.475 -8.475  8.475 /
\hshade -8.475  8.475  8.475 <z,,,>  0  8.475  12.0   8.475  8.475  8.475 /
\endpicture
\nonumber \\
&&\quad +
\beginpicture
\setcoordinatesystem units <1.00truept,1.00truept>
\setplotarea x from -60 to 40, y from -12 to 12
\circulararc 360 degrees from  12 0 center at 0 0
\circulararc 360 degrees from  34 0 center at 42 0
\circulararc 360 degrees from -22 0 center at -30 0
\plot  50 0   60 0 /
\plot  27 0   34 0 /
\plot  21 0   12 0 /
\plot -12 0  -22 0 /
\plot -38 0  -48 0 /
\put {{\tiny $1PI$}} [c] at  42 0
\put {{\tiny $1PI$}} [c] at -30 0
\put {{\scriptsize $\otimes$}} [c] at 24 0
\put {{\tiny $\vec x\cdot\vec\nabla_{\vec x}$}} [c] at 24 8
\put {{\tiny $t_2, y_2, s_2$}} [c] at -54 -7
\put {{\tiny $t_3, y_3, s_3$}} [c] at  66 -7
\setshadesymbol ({\tmrms .})
\setshadegrid span <0.9pt>
\setquadratic
\hshade -8.475 -8.475 -8.475 <,z,,>  0 -12.0   -8.475 8.475 -8.475 -8.475 /
\vshade -8.475 -8.475  8.475 <z,z,,> 0 -12.0   12.0   8.475 -8.475  8.475 /
\hshade -8.475  8.475  8.475 <z,,,>  0  8.475  12.0   8.475  8.475  8.475 /
\endpicture
\nonumber 
\end{eqnarray}
On the left side of this equation we are summing over all possible space-time points for the corresponding leg\footnote{Another note for the experts:  if we think of the times as defined on the full time-contour, the $\pm$ indices ($s_2$ and $s_3$) are naturally absorbed into the contour time-coordinate.}; on the right the $\vec x\cdot\vec\nabla_{\vec x}$ operator is acting on the connected two-point Green's function from both sides.

Since the connected and 1PI two-point functions are the functional inverses of each other, if we act on this equation with a pair of propagators, we amputate the 1PI legs on the right side and produce an expression on the left which can be written as a 1PI two-point function acting on the connected three-point function as we wanted,
\begin{eqnarray}
&&\!\!\!\!\!\!\!\!\!\!\!\!\!\!
-i\sum_{s=\pm}\int d^4y\, \int d^4x\, 
\Bigl\{ \Gamma^{+s}(y,x)\, G^{s++}_c(x,x_2,x_3) 
+ \Gamma^{-s}(y,x)\, G^{s++}_c(x,x_2,x_3) \Bigr\} 
\nonumber \\
&&\qquad\qquad\qquad\qquad\qquad\qquad\qquad\qquad\qquad\qquad
= \bigl[ \vec x_2\cdot\vec\nabla_{\vec x_2} 
+ \vec x_3\cdot\vec\nabla_{\vec x_3} \bigr] G^{++}_c(x_2,x_3) . 
\nonumber 
\end{eqnarray}
The full details for how to derive this equation appear in Appendix B.  Since for the Bunch-Davies state $\Gamma^{+-}=\Gamma^{-+}=0$, we have just 
\begin{eqnarray}
&&\!\!\!\!\!\!\!\!\!\!\!\!\!\!\!\!\!
-i \int d^4y\, \int d^4x\, 
\Bigl\{ \Gamma^{++}(y,x)\, G^{+++}_c(x,x_2,x_3) 
+ \Gamma^{--}(y,x)\, G^{-++}_c(x,x_2,x_3) \Bigr\} 
\nonumber \\
&&\qquad\qquad\qquad\qquad\qquad\qquad\qquad\qquad\qquad
= \bigl[ \vec x_2\cdot\vec\nabla_{\vec x_2} 
+ \vec x_3\cdot\vec\nabla_{\vec x_3} \bigr] G^{++}_c(x_2,x_3) . 
\nonumber 
\end{eqnarray}
This is the consistency relation between the three and two-point functions expressed in a fully time-evolving formalism.  The diagrammatic version of this identity may be written
$$
\beginpicture
\setcoordinatesystem units <1.00truept,1.00truept>
\setplotarea x from -72 to 36, y from -32 to 32
\circulararc 360 degrees from 12 0 center at 0 0
\circulararc 360 degrees from -36 0 center at -48 0
\plot -36 0  -12 0 /
\plot -84 0  -60 0 /
\plot  6  10.39  18  31.18 /
\plot  6 -10.39  18 -31.18 /
\put {{\footnotesize $1PI$}} [c] at -48 0
\put {$\displaystyle -i\int dtd^3\vec x\, $} [r] at -88 0
\put {{\scriptsize $t,\vec x$}} [l] at -84 -7
\put {{\scriptsize $t_*, x_2$}} [l] at 20  25
\put {{\scriptsize $t_*, x_3$}} [l] at 20 -28
\setshadesymbol ({\tmrms .})
\setshadegrid span <0.9pt>
\setquadratic
\hshade -8.475 -8.475 -8.475 <,z,,>  0 -12.0   -8.475 8.475 -8.475 -8.475 /
\vshade -8.475 -8.475  8.475 <z,z,,> 0 -12.0   12.0   8.475 -8.475  8.475 /
\hshade -8.475  8.475  8.475 <z,,,>  0  8.475  12.0   8.475  8.475  8.475 /
\endpicture
\beginpicture
\setcoordinatesystem units <1.00truept,1.00truept>
\setplotarea x from -90 to 40, y from -12 to 12
\circulararc 360 degrees from 12 0 center at 0 0
\plot -36 0  -12 0 /
\plot  12 0   36 0 /
\put {$=\ \ \bigl[ \vec x_2\cdot\vec\nabla_{\vec x_2} + \vec x_3\cdot\vec\nabla_{\vec x_3} \bigr]$} [r] at -48 0
\put {{\scriptsize $t_*, x_2$}} [c] at -36 -7
\put {{\scriptsize $t_*, x_3$}} [c] at  36 -7
\setshadesymbol ({\tmrms .})
\setshadegrid span <0.9pt>
\setquadratic
\hshade -8.475 -8.475 -8.475 <,z,,>  0 -12.0   -8.475 8.475 -8.475 -8.475 /
\vshade -8.475 -8.475  8.475 <z,z,,> 0 -12.0   12.0   8.475 -8.475  8.475 /
\hshade -8.475  8.475  8.475 <z,,,>  0  8.475  12.0   8.475  8.475  8.475 /
\endpicture
$$

This relation holds for the {\it full\/} Green's functions evaluated at any points and at any time during the inflationary era.  But to understand what it implies for the observations of our universe, we evaluate it at points on a late-time hypersurface at $t_*$.  Only two of the points in the relation are external, so we choose them lie somewhere on the this late-time hypersurface, 
$$
x_2 = (t_*,\vec x_2),\qquad
x_3 = (t_*,\vec x_3).
$$
$t_*$ is also meant to be the endpoint of the evolution, so it also appears as the upper limit of the time integrals, 
\begin{eqnarray}
&&\!\!\!\!\!\!\!\!\!\!\!
-i\int_{-\infty}^{t_*} dt'\, \int d^3\vec y\, 
\int_{-\infty}^{t_*} dt\, \int d^3\vec x\, \Bigl\{ 
\Gamma^{++}(y,x)\, G^{+++}_c(x,x_2,x_3) 
+ \Gamma^{--}(y,x)\, G^{-++}_c(x,x_2,x_3) 
\Bigr\}
\nonumber \\
&&\qquad\qquad\qquad\qquad\qquad\qquad\qquad\qquad\qquad\qquad\qquad
= \bigl[ \vec x_2\cdot\vec\nabla_{\vec x_2} 
+ \vec x_3\cdot\vec\nabla_{\vec x_3} \bigr]
G^{++}_c(x_2,x_3) ,
\nonumber 
\end{eqnarray}
where $y=(t',\vec y)$ and $x=(t,\vec x)$.  The relation is more conventionally expressed in terms of the spatial momenta, 
\begin{eqnarray}
&&\!\!\!\!\!\!\!\!\!\!\!
-i\int_{-\infty}^{t_*} dt'\, \int_{-\infty}^{t_*} dt\, 
\Bigl\{ 
\Gamma^{++}_0(t',t)\, G^{+++}_c(t,\vec 0;t_*,\vec k_2;t_*,\vec k_3) 
+ \Gamma^{--}_0(t',t)\, G^{-++}_c(t,\vec 0;t_*,\vec k_2;t_*,\vec k_3) 
\Bigr\}
\nonumber \\
&&\qquad\qquad\qquad\qquad\qquad\qquad\qquad\qquad\qquad\qquad\qquad
= -\bigl[ 3 + \vec k_2\cdot\nabla_{\vec k_2}\bigr]
G^{++}_c(t_*,\vec k_2;t_*,\vec k_3) . 
\nonumber 
\end{eqnarray}
While we have not written it, both sides are implicitly multiplied by a momentum-conserving $\delta^3(\vec k_2+\vec k_3)$.  The fact that we have integrated out one of the coordinates of the three-point function means that only the constant, $\vec k=\vec 0$, behaviour associated with that coordinate appears in the Fourier-transformed expression.

In this form, both sides of the consistency relation are finite.  On its left side, this property is a little more subtle, since the terms are products of something that diverges with something that vanishes in the limit
$$
\lim_{\vec k\to\vec 0} \bigl[ \Gamma^{\pm\pm}_k(t',t)\, 
G^{\pm ++}_c(t,\vec k;t_*,\vec k_2;t_*,\vec k_3) \bigr] \to \hbox{finite and nonzero.}
$$
What we really mean by the left side of the relation then is this limit.

The consistency relation is a statement about the full, `all orders in perturbation theory' Green's functions.  For the primordial fluctuations there is no need to go beyond the tree-level contributions---very often there is not even a need to go further than the leading expression in the slow-roll limit.  Only the amplitude and the leading information about the scale-dependence of the two-point function have been observed.  The three-point function has yet to be detected, though observations have constrained it to be quite small.  Therefore, using tree-level expressions for the 1PI two-point functions, found at the end of the last section, and integrating over a trivial $\delta$-function, we obtain our final expression for the consistency relation between the connected two and three-point functions, 
\begin{eqnarray}
&&\!\!\!\!\!\!\!\!\!\!\!\!\!\!\!\!\!\!\!\!
i \lim_{\vec k\to\vec 0} \int_{-\infty}^{t_*} dt\, 
\biggl\{ 
{\dot\phi^2\over\dot\rho^2} e^{3\rho} {d^2\over dt^2} 
+ {d\over dt}\biggl[ {\dot\phi^2\over\dot\rho^2} e^{3\rho} \biggr] {d\over dt}
+ k^2 {\dot\phi^2\over\dot\rho^2} e^\rho \biggr\}
\nonumber \\
&&\qquad\
\Bigl\{ 
G^{+++}_c(t,\vec k;t_*,\vec k_2;t_*,\vec k_3) 
- G^{-++}_c(t,\vec k;t_*,\vec k_2;t_*,\vec k_3) 
\Bigr\}
\nonumber \\
&&\qquad\qquad\qquad\qquad\qquad\qquad\qquad\qquad
= -\bigl[ 3 + \vec k_2\cdot\nabla_{\vec k_2}\bigr]
G^{++}_c(t_*,\vec k_2;t_*,\vec k_3) . 
\nonumber 
\end{eqnarray}
Once again, there is an implicit factor of $\delta^3(\vec k_2+\vec k_3)$ multiplying both sides.

Although it appears a little different from how the relation has often been written, the expression that we have found is in fact {\it formally\/} the same.  When the correlation functions are derived from a path integral effectively defined only on the late-time hypersurface, as was done in \cite{Goldberger:2013rsa}, the consistency relation assumes the form
$$
{G_c^{(3)}(\vec 0,\vec k_2,-\vec k_2)\over P(0)} 
= - \biggl[ 3 + k_2{\partial\over\partial k_2} \biggr] P(k_2) . 
$$
Here the Green's functions are the equal-time correlation functions evaluated at $t_*$; 
$$
G_c^{(3)}(\vec k_1,\vec k_2,\vec k_3) = G_c^{+++}(t_*,\vec k_1;t_*,\vec k_2;t_*,\vec k_3)  
$$
and $P(k)=G_c^{++}(t_*,\vec k;t_*,-\vec k)$ is the power spectrum.  If we recall that $[P(0)]^{-1}$ in that language is the 1PI two-point function evaluated in the zero-momentum limit, we see that this is structurally the same as the result above which has included the full time evolution.

It is nonetheless an instructive calculation to show how this relation applies to the simple class of slow-roll, single-field, models of inflation that Maldacena originally considered \cite{Maldacena:2002vr}.  This calculation is presented in Appendix A.  It has been provided in full because---as was mentioned---the left side contains a delicate, but finite, balance between diverging and vanishing factors.  We have also included it because some of the techniques, though they are becoming more commonly used in inflation, are still perhaps not as familiar as they ought to be.

\section{The Slavnov-Taylor identity with an initial state}

The advantage of following the full time evolution of the Green's functions is that it allows us to treat more general initial states.  We are no longer shackled to the Bunch-Davies state, or even to a particular class of symmetric initial states.  

Let us consider a universe where the scalar fluctuations $\zeta(t,\vec x)$ begin in an arbitrary state at $t=t_0$.  The information about this initial state can be incorporated into the generating functional by including a density matrix $\rho(t_0)$ for that state \cite{Agarwal:2012mq},
$$
Z[J^\pm] = \int {\cal D}\zeta^+\,{\cal D}\zeta^-\, \rho(t_0)\exp\Bigl\{ 
iS[\zeta^+]-iS[\zeta^-] + i \int_{t_0}^{t_*} dt\, \int d^3\vec x\, 
\bigl[ J^+\zeta^+ - J^-\zeta^- \bigr] \Bigr\} . 
$$
We write this matrix as an action defined along the initial-time hypersurface at $t_0$, 
$$
\rho(t_0) = e^{iS_0[\zeta^+(t_0,\vec x),\zeta^-(t_0,\vec x)]} ;
$$
this trick allows us to put it together with the rest of the action,
$$
Z[J^\pm] = \int {\cal D}\zeta^+\,{\cal D}\zeta^-\, \exp\Bigl\{ 
iS[\zeta^+]-iS[\zeta^-]+iS_0[\zeta^+,\zeta^-]
+ i \int_{t_0}^{t_*} dt\, \int d^3\vec x\, 
\bigl[ J^+\zeta^+ - J^-\zeta^- \bigr] \Bigr\}. 
$$
Each part of this action has its own meaning in the interaction picture:  the quadratic terms in $S$ define the time-dependence of the fields in the free theory, the higher-order interactions in $S$ determine the evolution of the state, and the terms in $S_0$ determine the initial state.  Whether we decide to group the quadratic terms of $S_0$ with those of $S$ when solving for the time-dependence of the free theory depends on the particular state.  When the initial state differs substantially from the Bunch-Davies state, so that a perturbative expansion provides a poor approximation, it is necessary to include all quadratic terms in the free part; but in cosmological examples, where the state does not appear to differ too much in its two-point structure from the Bunch-Davies state, it can sometimes be more convenient to regard all of $S_0$ as a part of the interactions.

The initial action is arranged as a series of terms according to powers of the fluctuations,\footnote{The linear term in $\zeta^\pm$ is included when it is necessary to cancel tadpole graphs involving the initial time.  We still are imposing the condition that the expectation value of $\zeta^\pm$ vanishes at all times.} 
$$
S_0[\zeta^\pm] = S_0^{(1)}[\zeta^\pm] + S_0^{(2)}[\zeta^\pm] 
+ S_0^{(3)}[\zeta^\pm] + S_0^{(4)}[\zeta^\pm] + \cdots, 
$$
where $S_0^{(2)}$ is quadratic in $\zeta^\pm$, $S_0^{(3)}$ is cubic, {\it etc.\/}  As long as the higher-order terms in this series are sufficiently small, their effects can be treated perturbatively.  Since the observed universe appears to be consistent with having a rather small non-Gaussian primordial component, we usually assume that we are within this regime.  The requirement that the initial density matrix is real, $\rho(t_0)=\rho^\dagger(t_0)$, imposes constraints on the form of $S_0$.  For example, when the state is translationally and rotationally invariant, the quadratic terms without time-derivatives of the field are specified by two functions, 
\begin{eqnarray}
S_0^{(2)}[\zeta^\pm] &=& 
- {1\over 2} \int d^3\vec x\, d^3\vec y\, \Bigl\{ 
\zeta^+(t_0,\vec x) A(\vec x-\vec y) \zeta^+(t_0,\vec y)
- \zeta^-(t_0,\vec x) A^*(\vec x-\vec y) \zeta^-(t_0,\vec y)
\nonumber \\
&&\qquad\qquad\quad\
+\,\, \zeta^+(t_0,\vec x) iB(\vec x-\vec y) \zeta^-(t_0,\vec y)
+ \zeta^-(t_0,\vec x) iB(\vec x-\vec y) \zeta^+(t_0,\vec y)
\Bigr\}
\nonumber \\
&&
+\,\, \cdots , 
\nonumber 
\end{eqnarray}
where $A(\vec x-\vec y)$ can be complex but $B(\vec x-\vec y)$ must be real.  The `$\cdots$' refer to further terms that could contain time derivatives of $\zeta$.  Similarly, the cubic terms, subject to the same assumptions, are fixed by two complex functions,
\begin{eqnarray}
S_0^{(3)}[\zeta^\pm] &=& 
- {1\over 6} \int d^3\vec x\, d^3\vec y\, d^3\vec z\, \Bigl\{ 
C(\vec x,\vec y,\vec z)\, \zeta^+(t_0,\vec x) \zeta^+(t_0,\vec y) \zeta^+(t_0,\vec z) 
\nonumber \\
&&\qquad\qquad\qquad\quad\!
-\,\, C^*(\vec x,\vec y,\vec z)\, \zeta^-(t_0,\vec x) \zeta^-(t_0,\vec y) \zeta^-(t_0,\vec z) 
\nonumber \\
&&\qquad\qquad\qquad\quad\!
+\,\, 3D(\vec x,\vec y,\vec z)\, \zeta^+(t_0,\vec x) \zeta^+(t_0,\vec y) \zeta^-(t_0,\vec z) 
\nonumber \\
&&\qquad\qquad\qquad\quad\!
-\,\, 3D^*(\vec x,\vec y,\vec z)\, \zeta^+(t_0,\vec x) \zeta^-(t_0,\vec y) \zeta^-(t_0,\vec z) 
+ \cdots \Bigr\} , 
\nonumber 
\end{eqnarray}
In writing the cubic surface action in this form we have tacitly assumed that the functions $C(\vec x,\vec y,\vec z)$ and $D(\vec x,\vec y,\vec z)$ are invariant under permutations of their coordinates.  Again, the cubic action could contain further operators with $\dot\zeta^\pm(t_0,\vec x)$ and higher derivatives.  Indeed, in the course of renormalizing the effects of the standard cubic operators that occur in inflation, operators with time derivatives do occur as counterterms in the initial action \cite{Collins:2014qna}.

Once we have included operators that contain odd numbers of the field, radiative corrections will typically produce tadpole graphs.  If they are to be cancelled, there must be linear terms in the action for the initial state,
$$
S_0^{(1)}[\zeta^\pm] = 
- \int d^3\vec x\, \Bigl\{ T(\vec x) \zeta^+(t_0,\vec x)
- T^*(\vec x) \zeta^-(t_0,\vec x) \Bigr\} + \cdots . 
$$
When the initial state is translationally and rotationally invariant, we expect that $T(\vec x)=T$ is a constant; but we shall leave it in this more general form for now.

Typically we shall find it more convenient to express the functions describing the state through their Fourier transforms,
$$
A(\vec x-\vec y) = \int {d^3\vec k\over (2\pi)^3}\, 
e^{i\vec k\cdot(\vec x-\vec y)} A_k ,
$$
and 
$$
C(\vec x,\vec y,\vec z) = \int {d^3\vec k_1\over (2\pi)^3} 
{d^3\vec k_2\over (2\pi)^3} {d^3\vec k_3\over (2\pi)^3}\, 
e^{i(\vec k_1\cdot\vec x+\vec k_2\cdot\vec y+\vec k_3\cdot\vec z)}\,
(2\pi)^3\, \delta^3(\vec k_1+\vec k_2+\vec k_3)\, 
C_{\vec k_1,\vec k_2,\vec k_3} ,
$$
for example.  Other than requiring that the higher energy modes are not so abundantly populated that they overwhelm the energy density of the inflationary background, these structures can have a more or less arbitrary dependence on the momentum.  

What sorts of initial states might we consider?  The initial time could be viewed in a variety of ways:  it could be simply a theoretical crutch, a cut-off that we impose to avoid considering the very early and correspondingly very short-distance behaviour of the theory.  On the other hand, it could genuinely be seen as a time when `something' happened---the beginning of the inflationary expansion or the end of some other dynamics still within inflation but beyond the minimal inflationary picture.  In the latter cases the moments before $t_0$ might have bequeathed the inflationary era with a state that was different from the standard Bunch-Davies state.  If we are taking a completely unprejudiced view of what happened before inflation, it is not {\it necessary\/} to assume that the initial state has the same symmetries of the single-field inflationary picture.  For computational convenience we shall still restrict to an initial state that is invariant under spatial translations and rotations.  But it is possible to allow initial states that break the spatial conformal symmetries.  These broken symmetries lead to interesting modifications of the standard consistency relations that we derived in the earlier sections.  In some instances too the conformal symmetries might be preserved though the state is not the Bunch-Davies state.

A state that is not invariant under the residual conformal symmetry will introduce additional terms in the primitive Slavnov-Taylor identity.  As in the earlier derivation of this identity, suppose that we make an infinitesimal conformal transformation of the field $\zeta(t,\vec x)$,
$$
\zeta(t,\vec x) \to \tilde\zeta(t,\vec x) 
= \zeta(t,\vec x) + \delta\zeta(t,\vec x) .
$$
For the purpose of deriving the Slavnov-Taylor identity, it is sufficient if this transformation is a symmetry of the action $S[\tilde\zeta]=S[\zeta]$ and if the measure of the functional integral changes at most by a functional constant, which can be absorbed into the normalization of $Z[J^\pm]$.  Both of these properties hold for the minimal inflationary picture.  When we write the generating functional in terms of $\tilde\zeta=\zeta+\delta\zeta$, by expanding the quantities that are not invariant under the symmetry to linear order, we obtain the change in the generating functional
\begin{eqnarray}
Z[J^\pm] &=& 
\int {\cal D}\tilde\zeta^\pm\, 
e^{iS[\tilde\zeta^+]-iS[\tilde\zeta^-]+iS_0[\tilde\zeta^\pm]
+ i \int_{t_0}^{t_*}dt\, \int d^3\vec x\, [J^+\tilde\zeta^+ -J^-\tilde\zeta^-] }
\nonumber \\
&=& 
\int {\cal D}\zeta^\pm\, 
e^{iS[\zeta^+]-iS[\zeta^-]+iS_0[\tilde\zeta^\pm]
+ i \int_{t_0}^{t_*}dt\, \int d^3\vec x\, [J^+\tilde\zeta^+ -J^-\tilde\zeta^-] }
\nonumber \\
&=& 
Z[J^\pm] + \delta Z[J^\pm] , 
\nonumber 
\end{eqnarray}
where
$$
\delta Z[J^\pm] = i \int_{t_0}^{t_*}dt\, \int d^3\vec x\, \Bigl\{ 
J^+(t,\vec x)\langle\delta\zeta^+(t,\vec x)\rangle_{J^\pm} 
- J^-(t,\vec x)\langle\delta\zeta^-(t,\vec x)\rangle_{J^\pm} \Bigr\}
+ i\langle\delta S_0[\zeta^\pm]\rangle_{J^\pm} .
$$
$\delta S_0$ represents the linear part of $S_0[\tilde\zeta^\pm]-S_0[\zeta^\pm]$ when expanded in powers of $\delta\zeta^\pm$.  The result, 
$$
\delta Z[J^\pm]=0 ,
$$
is again the `ur-statement' of the Slavnov-Taylor identity; however, it now contains additional contributions from the initial state.  From this identity we generate relations amongst various Green's functions by differentiating it with respect to $\bar\zeta^\pm(t,\vec x)$ an arbitrary number of times.

The most important class of consistency relations for inflation are those generated by dilations of the spatial part of the metric.  In its infinitesimal form, a dilation changes $\zeta(t,\vec x)$ by
$$
\delta\zeta = \lambda \bigl[ 1 + \vec x\cdot\vec\nabla\zeta \bigr] .
$$
When this expression is put into $\delta Z$, and the one-point functions are written in terms of the generating function for the connected Green's functions, $iW[J^\pm] = \ln Z[J^\pm]$, the Slavnov-Taylor identity becomes 
\begin{eqnarray}
&&\!\!\!\!\!\!\!\!\!\!\!\!\!\!\!\!\!\!
\int_{t_0}^{t_*}dt\, \int d^3\vec x\, \biggl\{ 
J^+(t,\vec x) - J^-(t,\vec x) 
+ J^+(t,\vec x)\, \vec x\cdot\vec\nabla 
{\delta W[J^\pm]\over\delta J^+(t,\vec x)}
+ J^-(t,\vec x)\, \vec x\cdot\vec\nabla 
{\delta W[J^\pm]\over\delta J^-(t,\vec x)}
 \biggr\}
\nonumber \\
&&\qquad\qquad\qquad\qquad\qquad\qquad\qquad\qquad\qquad\qquad\qquad\quad = 
- {1\over\lambda} {1\over Z[J^\pm]}\langle\delta S_0[\zeta^\pm]\rangle_{J^\pm} .
\nonumber 
\end{eqnarray}
We can broadly divide the initial states into two classes.  We define the first class to include those states that share the same conformal invariance---or here, the invariance under dilations---as as the inflationary metric.  The second class, where $\delta S_0\not=0$, represents the most general case.

\subsection{Conformally invariant initial states} 

So far the form of the initial state $S_0[\zeta^\pm]$ has been left largely arbitrary.  One formally simple class of initial states are those that are invariant under the same residual conformal symmetry of the inflationary metric.  Requiring that the state be conformally invariant imposes conditions on how the $n$-point structures of the initial state change under a conformal transformation.

Under a dilation the scalar fluctuation transforms inhomogeneously, so the terms in $\delta S_0^{(n)}$ have either $n$ or $n-1$ factors of the field $\zeta^\pm$.  This means that if we wish to have an initial state that is invariant under dilations, the change in the $n^{\rm th}$ order structure function is determined in part by the original function plus the integral of some linear combination of the $(n+1)^{\rm st}$ structure functions.  For example, if under a dilation 
$$
A(\vec x-\vec y) \to \tilde A(\vec x-\vec y) 
= A(\vec x-\vec y) + \lambda\, \delta A(\vec x-\vec y) ,
$$
then $\delta S_0=0$ when 
$$
\delta A(\vec x-\vec y) 
= \bigl[ 6 + \vec x\cdot\vec\nabla_{\vec x} + \vec y\cdot\vec\nabla_{\vec y} \bigr]\, A(\vec x-\vec y) 
- \int d^3\vec z\, \bigl[ C(\vec x,\vec y,\vec z) + D(\vec x,\vec y,\vec z)
\bigr] .
$$
Similarly, an invariant state should also have  
$$
\delta B(\vec x-\vec y) 
= \bigl[ 6 + \vec x\cdot\vec\nabla_{\vec x} + \vec y\cdot\vec\nabla_{\vec y} \bigr]\, B(\vec x-\vec y) 
+ i\int d^3\vec z\, \bigl[ D(\vec x,\vec y,\vec z) - D^*(\vec x,\vec y,\vec z)
\bigr] .
$$
The parts of the variation that are linear in the field require
$$
\delta T(\vec x) = \bigl[ 3 + \vec x\cdot\vec\nabla_{\vec x} \bigr] T(\vec x)
- \int d^3\vec y\, \bigl[ A(\vec x-\vec y) + iB(\vec x-\vec y) \bigr] .
$$
However, if were a imagining that the linear term has only been include to cancel tadpole graphs, so that it vanishes as we turn off the cubic structures and cubic interactions, then we might wish to have the linear and quadratic contributions to this equation vanish separately,
$$
\delta T(\vec x) = \bigl[ 3 + \vec x\cdot\vec\nabla_{\vec x} \bigr] T(\vec x)
\qquad\hbox{and}\qquad 
\int d^3\vec y\, \bigl[ A(\vec x-\vec y) + iB(\vec x-\vec y) \bigr] = 0 .
$$
Finally, the zeroth order part of the variation of the initial action under a dilation vanishes when 
$$
\int d^3\vec x\, \bigl[ T(\vec x) - T^*(\vec x) \bigr] = 0 ,
$$
which can be easily satisfied for a real $T(\vec x)$.

The Fourier-transformed versions of these conditions are 
\begin{eqnarray}
\delta A_k &=&
\bigl[ 3 - \vec k\cdot\vec\nabla_{\vec k} \bigr]\, A_k 
- C_{-\vec k,\vec k,\vec 0} - D_{-\vec k,\vec k,\vec 0}
\nonumber \\
\delta B_k &=&
\bigl[ 3 - \vec k\cdot\vec\nabla_{\vec k} \bigr]\, B_k 
+ i \bigl[ D_{-\vec k,\vec k,\vec 0} - D^*_{-\vec k,\vec k,\vec 0} \bigr] 
\nonumber 
\end{eqnarray}
and 
$$
A_0+iB_0 = 0 ,
$$
together with 
$$
\delta T_{\vec k} = - \vec k\cdot\vec\nabla_{\vec k} T_{\vec k} .
$$
The condition $A_0+iB_0=0$ refers to the spatially invariant part of the initial two-point structure; usually we assume that in this long-distance limit the state matches with the state that we have chosen as our reference state, and with respect to which we are defining the excited state.  Therefore, it is natural to let $A_0=0$ and $B_0=0$ and we assume that this is so even when we consider states that break the residual conformal symmetry of the background.

For a conformally invariant initial state, the statement of the Slavnov-Taylor identity {\it looks\/} exactly as it did for the Bunch-Davies state,
$$
\int_{t_0}^{t_*}dt\, \int d^3\vec x\, \biggl\{ 
J^+(t,\vec x) - J^-(t,\vec x) 
+ J^+(t,\vec x)\, \vec x\cdot\vec\nabla 
{\delta W[J^\pm]\over\delta J^+(t,\vec x)}
+ J^-(t,\vec x)\, \vec x\cdot\vec\nabla 
{\delta W[J^\pm]\over\delta J^-(t,\vec x)}
 \biggr\}
= 0 .
$$
Taking functional derivatives with respect to $\bar\zeta^\pm$ produces exactly the same infinite tower of consistency relations between $n$ and $n+1$-point correlation functions of the fluctuations produced by inflation.  However, this appearance is deceptive.  The consistency relations {\it amongst\/} the Green's functions have not changed, but the Green's functions themselves will be different.\footnote{The Green's functions for a general initial state are presented in \cite{Agarwal:2012mq} and \cite{Collins:2013kqa}.}  They are no longer the Green's functions for the Bunch-Davies state but are rather those for the appropriate initial state.  Phenomenologically this can lead to the case where the standard consistency relations appears to be violated.  But this spurious violation has come about not because the wrong consistency relations are being used, but rather because the wrong $n$-point functions have been assumed.  This interesting case is explored in \cite{Berezhiani:2014kga}.

\subsection{Broken conformal invariance in the initial state} 

On the other hand, there is no reason that the initial state must itself be invariant under the residual conformal symmetries of inflation.  The dynamics prior to $t_0$ did not need to have the same symmetries as what happened after $t_0$.  This freedom opens up much richer families of possibilities for the initial state.  The degree to which observations are in accord or violate the standard consistency relations can then be used to constrain the possible $n$-point structures of the initial state and how they transform conformally.  This more general case will be treated fully in later work \cite{generalstate}, but it is useful to explain the outlines of the calculation here. 

An interesting thing happens when we try to consider states that are not invariant under the full conformal symmetry group.  When we differentiate the Slavnov-Taylor identity $n>1$ times with respect to $\bar\zeta^\pm$ we obtain a non-vanishing result on the left side, which resembles the standard consistency relations except that, as for the conformally invariant initial state, the correlation functions are not those of the Bunch-Davies state.  However, unlike the invariant state, the functional derivative of the right side of the generalised Slavnov-Taylor identity does not vanish.

When we differentiate the Slavnov-Taylor identity {\it exactly\/} once, we obtain a tadpole condition on the initial state, 
$$
- {\delta\over\delta\bar\zeta^+(t,\vec x)}
{\langle\delta S_0[\zeta^\pm]\rangle_{J^\pm}\over \lambda Z[J^\pm]} = 0 ,
$$
which can be preserved by choosing the one-point structure, $T(\vec x)$, appropriately.  Once the tadpole has been fixed, it does not appear in any of the higher consistency relations---more than one functional derivative of the tadpole term with respect to $\bar\zeta^\pm$ annihilates it completely.

Our purpose here has been to develop a formalism that follows the full time-evolution within the Slavnov-Taylor relation, which allows it to be applied to arbitrary initial states.  These applications will be treated systematically elsewhere, but the basic recipe is as follows:  (1) choose a set of initial state structures and (2) specify how they transform under a conformal transformation.  In practice, since the consistency relations will be far more complicated, we might wish to make a few reasonable assumptions about the sizes of the higher order correlation functions relative to the lower order ones to be able to neglect some of the terms in the relation.

The contribution from the variation of the initial state can be rather complicated, even for the simplest structures.  For example, the contribution from the quadratic terms of $S_0$ to the right side of this Slavnov-Taylor identity for a dilation is
\begin{eqnarray}
&&\!\!\!\!\!\!\!\!\!\!\!\!\!\!\!\!\!\!\!\!\!\!\!\!\!\!
\int_{t_0}^{t_*}dt\, \int d^3\vec x\, \biggl\{ 
J^+(t,\vec x) - J^-(t,\vec x) 
+ J^+(t,\vec x)\, \vec x\cdot\vec\nabla 
{\delta W[J^\pm]\over\delta J^+(t,\vec x)}
+ J^-(t,\vec x)\, \vec x\cdot\vec\nabla 
{\delta W[J^\pm]\over\delta J^-(t,\vec x)}
\biggr\}
\nonumber \\
&=& 
{1\over 2} \int d^3\vec x\, d^3\vec y\, \biggl\{ 
A(\vec x-\vec y) \biggl[ 
{\delta W[J^\pm]\over\delta J^+(x)} + {\delta W[J^\pm]\over\delta J^+(y)}
-i\bigl[ \vec x\cdot\vec\nabla_{\vec x} + \vec y\cdot\vec\nabla_{\vec y} \bigr] {\delta^2 W[J^\pm]\over\delta J^+(x)\delta J^+(y)}
\nonumber \\
&&\qquad\qquad\qquad\qquad\qquad 
+ \bigl[ \vec x\cdot\vec\nabla_{\vec x} + \vec y\cdot\vec\nabla_{\vec y} \bigr] {\delta W[J^\pm]\over\delta J^+(x)}{\delta W[J^\pm]\over\delta J^+(y)} \biggr]
\nonumber \\
&&\qquad\qquad\quad 
+ \delta A(\vec x-\vec y) \biggl[ 
-i{\delta^2 W[J^\pm]\over\delta J^+(x)\delta J^+(y)} 
+{\delta W[J^\pm]\over\delta J^+(x)}{\delta W[J^\pm]\over\delta J^+(y)} \biggr]
\nonumber \\
&&\qquad\qquad\quad 
+ A^*(\vec x-\vec y) \biggl[ 
{\delta W[J^\pm]\over\delta J^-(x)} + {\delta W[J^\pm]\over\delta J^-(y)}
+i\bigl[ \vec x\cdot\vec\nabla_{\vec x} + \vec y\cdot\vec\nabla_{\vec y} \bigr] {\delta^2 W[J^\pm]\over\delta J^-(x)\delta J^-(y)}
\nonumber \\
&&\qquad\qquad\qquad\qquad\qquad 
- \bigl[ \vec x\cdot\vec\nabla_{\vec x} + \vec y\cdot\vec\nabla_{\vec y} \bigr] {\delta W[J^\pm]\over\delta J^-(x)}{\delta W[J^\pm]\over\delta J^-(y)} \biggr]
\nonumber \\
&&\qquad\qquad\quad 
+ \delta A^*(\vec x-\vec y) \biggl[ 
i{\delta^2 W[J^\pm]\over\delta J^-(x)\delta J^-(y)} 
-{\delta W[J^\pm]\over\delta J^-(x)}{\delta W[J^\pm]\over\delta J^-(y)} \biggr]
\nonumber \\
&&\qquad\qquad\quad 
+ 2iB(\vec x-\vec y) \biggl[ 
{\delta W[J^\pm]\over\delta J^+(x)} - {\delta W[J^\pm]\over\delta J^-(y)}
+i\bigl[ \vec x\cdot\vec\nabla_{\vec x} + \vec y\cdot\vec\nabla_{\vec y} \bigr] {\delta^2 W[J^\pm]\over\delta J^+(x)\delta J^-(y)}
\nonumber \\
&&\qquad\qquad\qquad\qquad\qquad 
- \bigl[ \vec x\cdot\vec\nabla_{\vec x} + \vec y\cdot\vec\nabla_{\vec y} \bigr] {\delta W[J^\pm]\over\delta J^+(x)}{\delta W[J^\pm]\over\delta J^-(y)} \biggr]
\nonumber \\
&&\qquad\qquad\quad 
+ 2i\delta B(\vec x-\vec y) \biggl[ 
i{\delta^2 W[J^\pm]\over\delta J^+(x)\delta J^-(y)} 
-{\delta W[J^\pm]\over\delta J^+(x)}{\delta W[J^\pm]\over\delta J^-(y)} \biggr]
+ \cdots \biggr\} ,
\nonumber 
\end{eqnarray}
and there are, of course, contributions from the cubic operators in $S_0^{(3)}$, the quartic operators, {\it etc.\/}  Differentiating this identity twice with respect to $\bar\zeta^\pm$ yields a relation between two and three-point correlators on the left side and a number of new terms on the right.

\section{Concluding remarks} 

The symmetry that remains even after we have made a general choice for the quantum fluctuations about an inflationary background leads to relations amongst their correlation functions.  We have seen that these relations are in fact the Slavnov-Taylor identities associated with this residual symmetry.  Because the approach that we have presented here only relies on symmetries of the metric and on very general properties of any quantum field theory, our results apply equally generally and are largely independent of the detailed properties of the particular model.

Throughout our derivation we have kept the evolution of the state intact and explicit.  This allows us immediately to adapt our approach to inflationary theories that start in more or less arbitrary initial states.  Having this freedom allows us to treat more general situations in inflation.  It also frees us from the need to make assumptions about the behaviour of the universe at asymptotically distant times and at infinitesimally small scales.  Since we do not know what might have occurred before an inflationary expansion, and moreover since there are dangers inherent in extending quantum field theories in inflating backgrounds arbitrarily far back in time---our understanding of nature is inadequate beyond certain scales---having the ability to choose other initial states and having the freedom to start at a finite initial time allows us to explore richer sets of possibilities.  Even viewed more conservatively, the formalism we have developed here lets us parametrise by how much the state can differ from a purely Bunch-Davies state---and then constrain these departures experimentally---rather than assume that the universe is in this state {\it ab initio\/}.  The same analysis equally applies to the tensor fluctuations, about whose state far less is known.

A general initial state modifies a consistency relation in two ways.  The Green's functions within the relation change to reflect the influence of the initial density matrix on observables, and the non-invariance of the initial state also alters the basic form of the Slavnov-Taylor identity from which the consistency relations are derived.  Because the underlying dynamics are still invariant under the residual conformal symmetries in the metric, it is still possible to derive a Slavnov-Taylor identity as long as we have included the appropriate corrections generated by the change of the initial density matrix under the conformal symmetry transformations.  These new terms in the identity are the cosmological analogues of the corrections that appear in the Ward identities of gauge theories in the presence of explicit symmetry breaking.

In this article we have derived the basic effect of an initial state on the Slavnov-Taylor identity.  We shall explore the possible observational effects of having non-Bunch-Davies and non-invariant states more fully in \cite{generalstate} using the formalism that we have developed here.  Most importantly, it would be interesting to see what the known properties of the power spectrum, and the constraints on the  non-Gaussianties, are able to tell us about the state during inflation.

While we have concentrated here on the consistency relation between the two and three-point functions for the scalar field---mainly to illustrate a new method with a familiar example---there are many further quantities to compute.  By differentiating the `ur-form' of the Slavnov-Taylor identity, $\delta Z[J^\pm]=0$, $n$ times with respect to $\bar\zeta^\pm(t,\vec x)$, we generate higher-order consistency relations between $n$ and $n+1$-point correlators.  Most of these are perhaps of a more formal interest, since thus far it has been difficult even to detect the amplitude of the three-point correlator.  Initial states could easily enhance some of these higher-order correlators without disturbing what we already know about the power spectrum.  Additionally, it would be interesting to study the consistency relation associated with the special conformal transformations.  

In his original derivation of the non-Gaussianities in inflation, Maldacena found further consistency relations satisfied by three-point functions which contained any combination of the tensor and scalar fluctuations.  This formalism can also be applied to such correlators.

\acknowledgments

R.~H.~and T.~V.~are grateful for the support of the Department of Energy (DE-FG03-91-ER40682) and of a grant from the John Templeton Foundation.  We would also like to thank Nishant Agarwal and Raquel Ribeiro for valuable conversations.

\appendix
\section{Evaluating the consistency relation} 

In this appendix we show that the standard slow-roll inflationary models with a single scalar inflaton field satisfy the consistency relation in the form that we have derived it, giving a---by now---standard estimate of the amplitude and scaling of the three-point function in the limit where one of the external momenta is soft.  While this conclusion is nothing new in itself, there are many technical points in treating and using the 1PI Green's functions which need to be thoroughly understood.

Before beginning this analysis, we mention a few of these points:  (1) Maldacena's trick of shifting the field in order to remove certain terms from the cubic action is not convenient here.  It would also entail a shift in the 1PI Green's function which is not so easy to do.  Fortunately, the leading parts of the action are simple enough to evaluate without the need to make this shift.  (2) In the limit in which the momentum of one of the external legs gets soft, the three-point function diverges.  But when we act upon it with a 1PI two-point function, we obtain a finite result.  At tree-level, the 1PI two-point function is 
$$
\Gamma^{++}_k(t',t)= - \Gamma^{--}_k(t',t) = - \delta(t'-t) \biggl\{ 
{\dot\phi^2\over\dot\rho^2} e^{3\rho} {d^2\over dt^2} 
+ {d\over dt}\biggl[ {\dot\phi^2\over\dot\rho^2} e^{3\rho} \biggr] {d\over dt}
+ k^2 {\dot\phi^2\over\dot\rho^2} e^\rho \biggr\} . 
$$
In the Bunch-Davies state the parts containing the time-derivatives {\it vanish\/} in the late-time limit giving a vanishing result when acting on the three-point function, for reasons that we shall explain.  We also should set the $k^2$ in the spatial derivative term {\it only after\/} this operator has acted on the three-point functions.
\vskip12truept

\subsection{Determining the leading operators} 

When the inflationary action is expanded to third order in the fluctuation $\zeta(t,\vec x)$, the resulting set of operators does not appear to be manifestly second-order in the slow-roll parameters:  a fair amount of further effort \cite{Collins:2011mz} is required to make this property self-evident.  Nearly every operator must be integrated by parts---often multiple times---with respect to its spatial or time derivatives, certain relations of the classical background must be imposed, {\it etc\/}.  At the end of this lengthy process, the cubic part of the action assumes the following form,
\begin{eqnarray}
S^{(3)} &\!\!\!=\!\!\!& 
\int d^4x\, \biggl\{
{1\over 4} {e^{3\rho}\over M_{\rm pl}^2} 
{\dot\phi^4\over\dot\rho^4} \dot\zeta^2 \zeta 
+ {1\over 4} {e^\rho\over M_{\rm pl}^2} {\dot\phi^4\over\dot\rho^4} 
\zeta \partial_k\zeta \partial^k\zeta 
- {1\over 2} {e^{3\rho}\over M_{\rm pl}^2} 
{\dot\phi^4\over\dot\rho^4} \dot\zeta \partial_k\zeta \partial^k\bigl(\partial^{-2}\dot\zeta \bigr)
\nonumber \\
&&\qquad\quad
+ {1\over 2}e^{3\rho} {\dot\phi^2\over\dot\rho^2} \dot\zeta \zeta^2 
{d\over dt} \biggl[ {\ddot\phi\over\dot\phi\dot\rho} 
+ {1\over 2M_{\rm pl}^2} {\dot\phi^2\over\dot\rho^2} \biggr] 
- {1\over 16} {e^{3\rho}\over M_{\rm pl}^2} {\dot\phi^6\over\dot\rho^6} 
\dot\zeta^2 \zeta
\nonumber \\
&&\qquad\quad
+ {1\over 16} {e^{3\rho}\over M_{\rm pl}^4} {\dot\phi^6\over\dot\rho^6} 
\zeta \partial_k\partial_l\bigl(\partial^{-2}\dot\zeta \bigr) 
\partial^k\partial^l\bigl(\partial^{-2}\dot\zeta \bigr) 
\nonumber \\
&&\qquad\quad
+ {1\over 2} \biggl\{ 
{d\over dt} \biggl[ e^{3\rho} {\dot\phi^2\over\dot\rho^2} \dot\zeta \biggr] 
- e^\rho {\dot\phi^2\over\dot\rho^2} \partial_k\partial^k\zeta 
\biggr\}
\nonumber \\
&&\qquad\qquad\times
\biggl\{ 
\biggl[ {\ddot\phi\over\dot\phi\dot\rho} 
+ {1\over 2} {1\over M_{\rm pl}^2} {\dot\phi^2\over\dot\rho^2} \biggr] \zeta^2 
+ 2 {1\over\dot\rho} \dot\zeta \zeta 
- {1\over 2} {e^{-2\rho}\over\dot\rho^2} \bigl[ \partial_k\zeta \partial^k\zeta 
- \partial^{-2} \partial_k\partial_l \bigl( \partial^k\zeta \partial^l\zeta \bigr) \bigr] 
\nonumber \\
&&\qquad\qquad\quad
+\,\, 
{1\over 2} {1\over\dot\rho} {1\over M_{\rm pl}^2} 
{\dot\phi^2\over\dot\rho^2} \bigl[ 
\partial_k \zeta \partial^k\bigl(\partial^{-2}\dot\zeta \bigr) 
- \partial^{-2} \partial_k\partial_l \bigl( \partial^k\zeta \partial^l\bigl(\partial^{-2}\dot\zeta \bigr) \bigr) \bigr] 
\biggr\} \biggr\} . 
\nonumber
\end{eqnarray}
This is the form of the cubic interactions that was derived by Maldacena.  Each of its operators either is manifestly second order or higher in the slow-roll parameters, 
$$
\epsilon = {1\over 2 M_{\rm pl}^2} {\dot\phi^2\over\dot\rho^2} 
\qquad\hbox{and}\qquad 
\delta = {\ddot\phi\over\dot\phi\dot\rho} ,
$$
or is proportional to the equation of motion for the quadratic part of the action.  A few terms even have both of these properties.  

It might seem that we have not quite succeeded in our goal, since not all the terms that are proportional to the equation of motion have enough factors of $\epsilon$ or $\delta$.  But by performing a nonlinear shift in the field, $\zeta\to\zeta_n+f(\zeta_n)$, where $f(\zeta_n)$ is quadratic in the field and is chosen precisely to remove the terms proportional to the equation of motion, we are left with just the first two lines of $S^{(3)}$.  Of course such a shift means that the three-point function of $\zeta$ will in turn be replaced by a sum of three and four-point functions of $\zeta_n$, which need to be separately computed.  In principle there are higher-point functions of $\zeta_n$ that appear too, but they are suppressed in the limits that we are considering.  Computing the three-point function in the late-time limit, the only part from these terms that contributes is 
$$
{1\over 2} \biggl\{ 
{d\over dt} \biggl[ e^{3\rho} {\dot\phi^2\over\dot\rho^2} \dot\zeta \biggr] 
- e^\rho {\dot\phi^2\over\dot\rho^2} \partial_k\partial^k\zeta 
\biggr\}
\biggl[ {\ddot\phi\over\dot\phi\dot\rho} 
+ {1\over 2} {\dot\phi^2\over\dot\rho^2} \biggr] \zeta^2 . 
$$
The shift removes it from the three-point function of $\zeta_n(t,\vec x)$, but it reappears through the four-point function of this $\zeta_n(t,\vec x)$ field.

This method for computing the three-point function of the fluctuation $\zeta(t,\vec x)$ is not especially convenient here.  The 1PI Green's functions that appear in the consistency relations are those associated with the original field $\zeta(t,\vec x)$ and not the shifted field $\zeta_n(t,\vec x)$.  We could try to figure out how correspondingly to alter the 1PI Green's functions so that they are compatible with the Green's functions for $\zeta_n(t,\vec x)$, but it is simpler to avoid the shift altogether.  In fact, the piece that contributes in the late-time limit, 
$$
{1\over 2} \biggl\{ 
{d\over dt} \biggl[ e^{3\rho} {\dot\phi^2\over\dot\rho^2} \dot\zeta \biggr] 
- e^\rho {\dot\phi^2\over\dot\rho^2} \partial_k\partial^k\zeta 
\biggr\}
\biggl[ {\ddot\phi\over\dot\phi\dot\rho} 
+ {1\over 2} {1\over M_{\rm pl}^2} {\dot\phi^2\over\dot\rho^2} \biggr] \zeta^2 , 
$$
is itself a linear combination of operators that are present in the first lines of the action as we wrote it earlier.  Integrating a time derivative by parts in the first term, and a spatial derivative in the second, yields 
\begin{eqnarray}
&&\!\!\!\!\!\!\!\!\!\!\!\!\!\!\!\!\!\!\!\!\!\!\!\!
{1\over 2} 
{d\over dt} \biggl[ e^{3\rho} {\dot\phi^2\over\dot\rho^2} \dot\zeta \biggr] 
\biggl[ {\ddot\phi\over\dot\phi\dot\rho} 
+ {1\over 2} {1\over M_{\rm pl}^2} {\dot\phi^2\over\dot\rho^2} \biggr] \zeta^2 
- {1\over 2} e^\rho {\dot\phi^2\over\dot\rho^2} \partial_k\partial^k\zeta 
\biggl[ {\ddot\phi\over\dot\phi\dot\rho} 
+ {1\over 2} {1\over M_{\rm pl}^2} {\dot\phi^2\over\dot\rho^2} \biggr] \zeta^2 
\nonumber \\
&=& 
-\,\, e^{3\rho} {\dot\phi^2\over\dot\rho^2} 
\biggl[ {\ddot\phi\over\dot\phi\dot\rho} 
+ {1\over 2} {1\over M_{\rm pl}^2} {\dot\phi^2\over\dot\rho^2} \biggr] \dot\zeta^2\zeta 
+ e^\rho {\dot\phi^2\over\dot\rho^2} 
\biggl[ {\ddot\phi\over\dot\phi\dot\rho} 
+ {1\over 2} {1\over M_{\rm pl}^2} {\dot\phi^2\over\dot\rho^2} \biggr] \zeta\partial_k\zeta\partial^k\zeta 
\nonumber \\
&& 
-\,\, {1\over 2} e^{3\rho} {\dot\phi^2\over\dot\rho^2} \dot\zeta \zeta^2 
{d\over dt} \biggl[ {\ddot\phi\over\dot\phi\dot\rho} 
+ {1\over 2} {1\over M_{\rm pl}^2} {\dot\phi^2\over\dot\rho^2} \biggr] 
+ \hbox{total derivatives} .
\nonumber 
\end{eqnarray}
Combining their coefficients with those of the first two operators in the cubic action, the set of operators that simultaneously are second-order in $\epsilon$ and $\delta$ and have contributions that do not vanish in the late-time limit reduces to just three operators, 
\begin{eqnarray}
S^{(3)} &\!\!\!=\!\!\!& 
\int d^4x\, \biggl\{
- e^{3\rho} {\dot\phi^2\over\dot\rho^2} 
\biggl[ {\ddot\phi\over\dot\phi\dot\rho} 
+ {1\over 4} {1\over M_{\rm pl}^2} {\dot\phi^2\over\dot\rho^2} \biggr] \dot\zeta^2\zeta 
+ e^\rho {\dot\phi^2\over\dot\rho^2} 
\biggl[ {\ddot\phi\over\dot\phi\dot\rho} 
+ {3\over 4} {1\over M_{\rm pl}^2} {\dot\phi^2\over\dot\rho^2} \biggr] \zeta\partial_k\zeta\partial^k\zeta 
\nonumber \\
&&\qquad\quad
- {1\over 2} {e^{3\rho}\over M_{\rm pl}^2}{\dot\phi^4\over\dot\rho^4} 
\dot\zeta \partial_k\zeta \partial^k\bigl(\partial^{-2}\dot\zeta \bigr)
+ \cdots \biggr\} . 
\nonumber 
\end{eqnarray}
Thus, to find the difference of the three-point functions that appear in the consistency relation requires computing the contributions of these three operators.  To make some of the intermediate stages of the following calculation a little less cluttered, let us abbreviate these coefficients of the operators by 
\begin{eqnarray}
\alpha\, M_{\rm pl}^2 &=& 
- {\dot\phi^2\over\dot\rho^2} 
\biggl[ {\ddot\phi\over\dot\phi\dot\rho} 
+ {1\over 4} {1\over M_{\rm pl}^2} {\dot\phi^2\over\dot\rho^2} \biggr] 
= - \epsilon \bigl[ \epsilon + 2\delta \bigr] \, M_{\rm pl}^2
\nonumber \\
\beta\, M_{\rm pl}^2 &=& 
{\dot\phi^2\over\dot\rho^2} 
\biggl[ {\ddot\phi\over\dot\phi\dot\rho} 
+ {3\over 4} {1\over M_{\rm pl}^2} {\dot\phi^2\over\dot\rho^2} \biggr]
= \epsilon \bigl[ 3\epsilon + 2\delta \bigr] \, M_{\rm pl}^2
\nonumber \\
\gamma\, M_{\rm pl}^2 &=& 
- {1\over 2} {1\over M_{\rm pl}^2} {\dot\phi^4\over\dot\rho^4} 
= - 2\epsilon^2 \, M_{\rm pl}^2 ,
\nonumber 
\end{eqnarray}
so that they might be written as 
$$
{\cal O}_1 = \alpha\, M_{\rm pl}^2\, e^{3\rho} \dot\zeta^2\zeta ,
\qquad
{\cal O}_2 = \beta\, M_{\rm pl}^2\, e^\rho \zeta\partial_k\zeta\partial^k\zeta ,
\qquad
{\cal O}_3 = \gamma\, M_{\rm pl}^2\, e^{3\rho} \dot\zeta \partial_k\zeta \partial^k\bigl(\partial^{-2}\dot\zeta \bigr) . 
$$
Defined in this way, $\alpha$, $\beta$, and $\gamma$ are dimensionless parameters which are all quadratic in the slow-roll parameters.

\subsection{A simplification and a subtlety} 

Having found what we need to compute, let us next determine what each of these operators does in fact contribute to the difference, 
$$
G_c^{+++}(t,\vec k;t_*,\vec k_2;t_*,\vec k_3) 
- G_c^{-++}(t,\vec k;t_*,\vec k_2;t_*,\vec k_3) ,
$$
which will be evaluated in the limit where $\vec k\to\vec 0$, $\vec k_3\to -\vec k_2$.  The definitions of these Green's functions were explained earlier.  Using a perturbative expansion organised in powers of the interaction Hamiltonian, which in this case is 
$$
H_I(t') = - \int d^3\vec w\, \Bigl\{ {\cal O}_1(t',\vec w) 
+ {\cal O}_2(t',\vec w) + {\cal O}_3(t',\vec w) + \cdots \Bigr\} , 
$$
the leading contribution to the difference comes from the terms with just one power of $H_I(t')$---or equivalently, just one power of each operator.  We shall define the contribution from each operator by writing a corresponding subscript on the three-point function, 
\begin{eqnarray}
&&\!\!\!\!\!\!\!\!\!\!\!\!\!\!\!\!
G_i^{+++}(t,\vec x;t_*,\vec x_2;t_*,\vec x_3) 
- G_i^{-++}(t,\vec x;t_*,\vec x_2;t_*,\vec x_3) 
\nonumber \\
&=& 
i\int_{-\infty}^{t_*} dt' \int d^3\vec w\, 
\langle 0|T\bigl(\zeta^+(t,\vec x)\zeta^+(t_*,\vec x_2)\zeta^+(t_*,\vec x_3) \bigl[ {\cal O}_i^+(t',\vec w) - {\cal O}_i^-(t',\vec w) \bigr] \bigr)|0\rangle
\nonumber \\
&& 
-\,\, i\int_{-\infty}^{t_*} dt' \int d^3\vec w\, 
\langle 0|T\bigl(\zeta^-(t,\vec x)\zeta^+(t_*,\vec x_2)\zeta^+(t_*,\vec x_3) \bigl[ {\cal O}_i^+(t',\vec w) - {\cal O}_i^-(t',\vec w) \bigr] \bigr)|0\rangle
+ \cdots
\nonumber \\
&=& 
i\int_{-\infty}^{t_*} dt' \int d^3\vec w\, \Bigl\{
\langle 0|T\bigl( \bigl[ \zeta^+(t,\vec x)- \zeta^-(t,\vec x) \bigr] \zeta^+(t_*,\vec x_2)\zeta^+(t_*,\vec x_3) {\cal O}_i^+(t',\vec w) \bigr)|0\rangle
\nonumber \\
&&\qquad\qquad\qquad
-\,\,  
\langle 0|T\bigl( \bigl[ \zeta^+(t,\vec x)- \zeta^-(t,\vec x) \bigr] \zeta^+(t_*,\vec x_2)\zeta^+(t_*,\vec x_3) {\cal O}_i^-(t',\vec w) \bigr)|0\rangle \Bigr\}
+ \cdots .
\nonumber 
\end{eqnarray}
Once we have performed all of the Wick contractions of the fields, we transform the result to momentum space and proceed to evaluate what it adds to the consistency relation.

The leading contribution of each operator to the left side of the consistency relation is then found by computing 
\begin{eqnarray}
&&\!\!\!\!\!\!\!\!\!\!\!\!\!\!
i\int_{-\infty}^{t_*} dt^{\prime\prime}\, \int_{-\infty}^{t_*} dt\, 
\delta(t^{\prime\prime}-t)\, \biggl\{ 
{\dot\phi^2\over\dot\rho^2} e^{3\rho(t)} {d^2\over dt^2} 
+ {d\over dt}\biggl[ {\dot\phi^2\over\dot\rho^2} e^{3\rho(t)} \biggr] {d\over dt}
+ k^2 {\dot\phi^2\over\dot\rho^2} e^{\rho(t)} \biggr\} 
\nonumber \\
&&\qquad\qquad\qquad\qquad\quad
\Bigl\{ 
G^{+++}_i(t,\vec k;t_*,\vec k_2;t_*,\vec k_3) 
- G^{-++}_i(t,\vec k;t_*,\vec k_2;t_*,\vec k_3) 
\Bigr\}
\nonumber \\
&&\qquad\quad
= i\int_{-\infty}^{t_*} dt\, \biggl\{ 
{\dot\phi^2\over\dot\rho^2} e^{3\rho(t)} {d^2\over dt^2} 
+ {d\over dt}\biggl[ {\dot\phi^2\over\dot\rho^2} e^{3\rho(t)} \biggr] {d\over dt}
+ k^2 {\dot\phi^2\over\dot\rho^2} e^{\rho(t)} \biggr\} 
\nonumber \\
&&\qquad\qquad\qquad\qquad
\Bigl\{ 
G^{+++}_i(t,\vec k;t_*,\vec k_2;t_*,\vec k_3) 
- G^{-++}_i(t,\vec k;t_*,\vec k_2;t_*,\vec k_3) 
\Bigr\}
\nonumber 
\end{eqnarray}
in the limit where $k\to 0$.  Before going further, however, we point out a simplification that occurs when the difference between the three-point functions has the structure,
$$
G^{+++}_i(t,\vec k;t_*,\vec k_2;t_*,\vec k_3) 
- G^{-++}_i(t,\vec k;t_*,\vec k_2;t_*,\vec k_3) 
= - \int_{-\infty}^{t_*} dt'\, \Theta(t'-t)\, F(t,t') ,
$$
which is true for each of the three operators, though the function $F(t,t')$ will be different for each.  Let us look at the part of the consistency relation that contains the time-derivatives from the 1PI Green's functions acting on something of this form, 
\begin{eqnarray}
&&\!\!\!\!\!\!\!\!\!\!\!\!\!\!
- i\int_{-\infty}^{t_*} dt\, \int_{-\infty}^{t_*} dt'\, 
\biggl\{ 
{\dot\phi^2\over\dot\rho^2} e^{3\rho(t)} {d^2\over dt^2} 
+ {d\over dt}\biggl[ {\dot\phi^2\over\dot\rho^2} e^{3\rho(t)}\biggr] {d\over dt}
\biggr\} 
\Bigl\{ \Theta(t'-t)\, F(t,t') \Bigr\}
\nonumber \\
&=&
- i\int_{-\infty}^{t_*} dt\, \int_{-\infty}^{t_*} dt'\, 
\biggl\{ 
- {\dot\phi^2\over\dot\rho^2} e^{3\rho(t)}  
\biggl[{d\over dt}\delta(t'-t) \biggr]\, F(t,t') 
- 2{\dot\phi^2\over\dot\rho^2} e^{3\rho(t)}  
\delta(t'-t)\, {d\over dt} F(t,t') 
\nonumber \\
&&\qquad\qquad\qquad\qquad
- {d\over dt}\biggl[ {\dot\phi^2\over\dot\rho^2} e^{3\rho(t)}\biggr] 
\delta(t'-t)\, F(t,t') 
+ \Theta(t'-t)\, {d\over dt}\biggl[ {\dot\phi^2\over\dot\rho^2} e^{3\rho(t)} {d\over dt}F(t,t') \biggr] 
\biggr\} . 
\nonumber 
\end{eqnarray}

The derivative of the $\delta$-function must be treated carefully.  If we {\it define\/} 
$$
\int_{t_1}^{t_2} dt\, \dot\delta(t-t_0)\, f(t) \equiv - \dot f(t_0) ,
$$
{\it without any boundary terms included\/}, then we obtain the correct result for the consistency relation.  Therefore we shall define $\dot\delta(t)$ by this relation, rather than attempt to subtract explicitly the unwanted boundary terms that would otherwise occur.  Applying this definition to the expression above yields, 
\begin{eqnarray}
&&\!\!\!\!\!\!\!\!\!\!\!\!
- i\int_{-\infty}^{t_*} dt\, \int_{-\infty}^{t_*} dt'\, 
\biggl\{ 
{\dot\phi^2\over\dot\rho^2} e^{3\rho(t)} {d^2\over dt^2} 
+ {d\over dt}\biggl[ {\dot\phi^2\over\dot\rho^2} e^{3\rho(t)}\biggr] {d\over dt}
\biggr\} 
\Bigl\{ \Theta(t'-t)\, F(t,t') \Bigr\}
\nonumber \\
&=&
- i\int_{-\infty}^{t_*} dt\, \int_{-\infty}^{t_*} dt'\, 
\biggl\{ 
- {\dot\phi^2\over\dot\rho^2} e^{3\rho(t)} \delta(t'-t)\, {d\over dt} F(t,t') 
+ \Theta(t'-t)\, {d\over dt}\biggl[ {\dot\phi^2\over\dot\rho^2} e^{3\rho(t)} {d\over dt}F(t,t') \biggr] 
\biggr\} 
\nonumber \\
&=&
i\int_{-\infty}^{t_*} dt'\, 
{\dot\phi^2\over\dot\rho^2} e^{3\rho(t')}\, \biggl[ {d\over dt} F(t,t') \biggr]_{t=t'}
- i\int_{-\infty}^{t_*} dt'\, \int_{-\infty}^{t'} dt\, 
\biggl\{ 
{d\over dt}\biggl[ {\dot\phi^2\over\dot\rho^2} e^{3\rho(t)} {d\over dt}F(t,t') \biggr] 
\biggr\} .
\nonumber 
\end{eqnarray}
The second term on the last line is a total derivative, so we can write it as 
$$
\int_{-\infty}^{t'} dt\, 
\biggl\{ 
{d\over dt}\biggl[ {\dot\phi^2\over\dot\rho^2} e^{3\rho(t)} {d\over dt}F(t,t') \biggr] \biggr\} 
= {\dot\phi^2\over\dot\rho^2} e^{3\rho(t')} \biggl[ {d\over dt}F(t,t') \biggr]_{t=t'} , 
$$
where we have assumed that there is no contribution from the $t\to -\infty$ boundary.\footnote{This relation will require an initial boundary term, however, when we start the system in a more general initial state at $t=t_0$.}  Therefore, the two terms exactly cancel, leaving no contribution from the time-derivative parts of the 1PI Green's functions, 
\begin{eqnarray}
&&\!\!\!\!\!\!\!\!\!\!\!\!\!\!\!\!\!\!\!\!
- i\int_{-\infty}^{t_*} dt\, \int_{-\infty}^{t_*} dt'\, 
\biggl\{ 
{\dot\phi^2\over\dot\rho^2} e^{3\rho(t)} {d^2\over dt^2} 
+ {d\over dt}\biggl[ {\dot\phi^2\over\dot\rho^2} e^{3\rho(t)}\biggr] {d\over dt}
\biggr\} 
\Bigl\{ \Theta(t'-t)\, F(t,t') \Bigr\}
\nonumber \\
&=&
i\int_{-\infty}^{t_*} dt'\, 
{\dot\phi^2\over\dot\rho^2} e^{3\rho(t')}\, \biggl[ {d\over dt} F(t,t') \biggr]_{t=t'}
- i\int_{-\infty}^{t_*} dt'\, {\dot\phi^2\over\dot\rho^2} e^{3\rho(t')} \biggl[ {d\over dt}F(t,t') \biggr]_{t=t'}
= 0 .
\nonumber 
\end{eqnarray}

As long as the difference of the three-point functions has the form that we assumed, then to leading order, the only part of the consistency relation that we need to calculate in detail, is the part due to the spatial derivatives, 
\begin{eqnarray}
&&\!\!\!\!\!\!\!\!\!\!\!\!\!\!\!\!\!\!\!\!
i\int_{-\infty}^{t_*} dt^{\prime\prime}\, \int_{-\infty}^{t_*} dt\, 
\delta(t^{\prime\prime}-t)\, \biggl\{ 
{\dot\phi^2\over\dot\rho^2} e^{3\rho(t)} {d^2\over dt^2} 
+ {d\over dt}\biggl[ {\dot\phi^2\over\dot\rho^2} e^{3\rho(t)} \biggr] {d\over dt}
+ k^2 {\dot\phi^2\over\dot\rho^2} e^{\rho(t)} \biggr\} 
\nonumber \\
&&\qquad\qquad\qquad\qquad
\Bigl\{ 
G^{+++}_i(t,\vec k;t_*,\vec k_2;t_*,\vec k_3) 
- G^{-++}_i(t,\vec k;t_*,\vec k_2;t_*,\vec k_3) 
\Bigr\}
\nonumber \\
&&\!\!\!\!\!\!\!\!\!\!\!\!\!\!\!\!\!\!\!\!
= \lim_{k\to 0} \biggl\{ ik^2 \int_{-\infty}^{t_*} dt\, 
{\dot\phi^2\over\dot\rho^2} e^{\rho(t)} 
\Bigl[ G^{+++}_i(t,\vec k;t_*,\vec k_2;t_*,\vec k_3) 
- G^{-++}_i(t,\vec k;t_*,\vec k_2;t_*,\vec k_3) \Bigr] \biggr\} . 
\nonumber 
\end{eqnarray}

\subsection{The second operator} 

We begin with the second operator since its structure is the simplest, having only spatial derivatives.  Substituting this operator into the above expression and taking the Wick contractions, what results is 
\begin{eqnarray}
&&\!\!\!\!\!\!\!\!\!\!\!\!\!\!\!\!
G_2^{+++}(t,\vec x;t_*,\vec x_2;t_*,\vec x_3) 
- G_2^{-++}(t,\vec x;t_*,\vec x_2;t_*,\vec x_3) 
\nonumber \\
&=& 
\int {d^3\vec k\over (2\pi)^3} \int {d^3\vec k_2\over (2\pi)^3}
\int {d^3\vec k_3\over (2\pi)^3}\,
e^{i\vec k\cdot\vec x} e^{i\vec k_2\cdot\vec x_2} e^{i\vec k_3\cdot\vec x_3}\, 
(2\pi)^3\, \delta^3(\vec k+\vec k_2+\vec k_3)
\nonumber \\
&&
i\beta\, M_{\rm pl}^2\,\bigl[ k^2+k_2^2+k_3^2 \bigr]\, 
\int_{-\infty}^{t_*} dt'\, e^{\rho(t')} \Bigl\{ 
[G_k^{++}(t,t')-G_k^{-+}(t,t')] G_{k_2}^{++}(t_*,t')G_{k_3}^{++}(t_*,t')
\nonumber \\
&&\qquad\qquad\qquad\qquad\qquad\qquad\quad\, 
-\,\,  
[G_k^{+-}(t,t')-G_k^{--}(t,t')] G_{k_2}^{+-}(t_*,t')G_{k_3}^{+-}(t_*,t')
\Bigr\} .
\nonumber 
\end{eqnarray}
To convert the Feynman propagators $G_{k_i}^{\pm\pm}$ into the appropriate time-ordered combinations of Wightman functions, notice that the $k_2$ and $k_3$-dependent factors are especially simple since $t_*$ always occurs later than $t'$.  For the $k$-dependent parts, we find that everything is proportional to a common $\Theta$-function,
\begin{eqnarray}
G_k^{++}(t,t')-G_k^{-+}(t,t')
&=& 
\Theta(t-t')\, G_k^>(t,t') + \Theta(t'-t)\, G_k^<(t,t') - G_k^>(t,t')
\nonumber \\
&=& 
-\Theta(t'-t)\, \bigl[ G_k^>(t,t') - G_k^<(t,t') \bigr] 
\nonumber \\
G_k^{+-}(t,t')-G_k^{--}(t,t')
&=& 
G_k^<(t,t') - \Theta(t'-t)\, G_k^>(t,t') - \Theta(t-t')\, G_k^<(t,t') 
\nonumber \\
&=& 
- \Theta(t'-t)\, \bigl[ G_k^>(t,t') - G_k^<(t,t') \bigr] . 
\nonumber 
\end{eqnarray}
The Fourier-transform of the contribution from the second operator to the difference in the three-point functions is then
\begin{eqnarray}
&&\!\!\!\!\!\!\!\!\!\!\!\!\!\!\!\!\!\!\!\!\!\!\!\!\!
G_2^{+++}(t,\vec k;t_*,\vec k_2;t_*,\vec k_3) 
- G_2^{-++}(t,\vec k;t_*,\vec k_2;t_*,\vec k_3) 
\nonumber \\
&=& 
-i\beta\, M_{\rm pl}^2\,\bigl[ k^2+k_2^2+k_3^2 \bigr]\, 
\int_{-\infty}^{t_*} dt'\, e^{\rho(t')} \Theta(t'-t)\, 
\bigl[ G_k^>(t,t') - G_k^<(t,t') \bigr]
\nonumber \\
&&\qquad\qquad\qquad\qquad\qquad
\bigl[ G_{k_2}^>(t_*,t')G_{k_3}^>(t_*,t') 
- G_{k_2}^<(t_*,t')G_{k_3}^<(t_*,t') \bigr] .
\nonumber 
\end{eqnarray}
We should mention here that this same structure, this combination of Wightman functions,
$$
\Theta(t'-t)\, \bigl[ G_k^>(t,t') - G_k^<(t,t') \bigr]
\bigl[ G_{k_2}^>(t_*,t')G_{k_3}^>(t_*,t') 
- G_{k_2}^<(t_*,t')G_{k_3}^<(t_*,t') \bigr] , 
$$
occurs in the calculations of the other two operators, though with additional time-derivatives and momentum factors.  It has exactly the $\Theta(t'-t)\, F(t,t')$ structure that is needed for the time-derivative parts of the 1PI Green's functions not to contribute.

The only contribution to the consistency relation for the Bunch-Davies state can come from the spatial derivative term, 
\begin{eqnarray}
&&\!\!\!\!\!\!\!\!\!\!
i k^2 \int_{-\infty}^{t_*} dt\, 
{\dot\phi^2\over\dot\rho^2} e^{\rho(t)} \Bigl\{ 
G_2^{+++}(t,\vec k;t_*,\vec k_2;t_*,\vec k_3) 
- G_2^{-++}(t,\vec k;t_*,\vec k_2;t_*,\vec k_3) \Bigr\} 
\nonumber \\
&=& 
i k^2 \int_{-\infty}^{t_*} dt\, 
{\dot\phi^2\over\dot\rho^2} e^{\rho(t)} \biggl\{ 
-i\beta\, M_{\rm pl}^2\,\bigl[ k^2+k_2^2+k_3^2 \bigr]\, 
\int_{-\infty}^{t_*} dt'\, e^{\rho(t')} \Theta(t'-t)\, 
\nonumber \\
&&\qquad\qquad\qquad
\bigl[ G_k^>(t,t') - G_k^<(t,t') \bigr]
\bigl[ G_{k_2}^>(t_*,t')G_{k_3}^>(t_*,t') 
- G_{k_2}^<(t_*,t')G_{k_3}^<(t_*,t') \bigr] \biggr\} 
\nonumber \\
&=& 
k^2 \beta\, M_{\rm pl}^2 
{\dot\phi^2\over\dot\rho^2}\,\bigl[ k^2+k_2^2+k_3^2 \bigr]\, 
\int_{-\infty}^{t_*} dt'\, e^{\rho(t')} 
\bigl[ G_{k_2}^>(t_*,t')G_{k_3}^>(t_*,t') 
- G_{k_2}^<(t_*,t')G_{k_3}^<(t_*,t') \bigr]
\nonumber \\
&&\qquad\qquad\qquad\qquad\qquad\times
\int_{-\infty}^{t'} dt\, e^{\rho(t)} 
\bigl[ G_k^>(t,t') - G_k^<(t,t') \bigr] . 
\nonumber 
\end{eqnarray}

Let us evaluate these two nested time-integrals.  As we are only working to leading order in the slow-roll parameters, we can write the scale factor and the Wightman functions in the de Sitter limit.  These functions assume a simpler form when expressed in terms of the conformal time  $(\eta,\eta', \eta_*)$ rather than the time coordinates that we have been using, $(t,t', t_*)$.  For example, the measure and scale factor become
$$
dt\, e^{\rho(t)} = d\eta\, {dt\over d\eta} e^{\rho(t)}
= d\eta\, {1\over -H\eta}{1\over -H\eta} 
= {1\over H^2} {d\eta\over \eta^2} . 
$$
The Wightman functions in the de Sitter limit are given by
\begin{eqnarray}
G_k^>(t,t') &=& 
{1\over 4\epsilon} {H^2\over M_{\rm pl}^2} {1\over k^3} (1+ik\eta)(1-ik\eta')
e^{-ik(\eta-\eta')} + \cdots 
\nonumber \\
G_k^<(t,t') &=& 
{1\over 4\epsilon} {H^2\over M_{\rm pl}^2} {1\over k^3} (1-ik\eta)(1+ik\eta')
e^{ik(\eta-\eta')} + \cdots , 
\nonumber 
\end{eqnarray}
where we do need to retain the initial $1/\epsilon$ factor since it would otherwise diverge in the strict de Sitter limit.  The $dt$-integral then becomes 
\begin{eqnarray}
&&\!\!\!\!\!\!\!\!\!\!\!\!\!\!\!\!
\int_{-\infty}^{t'} dt\, e^{\rho(t)} \bigl[ G_k^>(t,t') - G_k^<(t,t') \bigr] 
\nonumber \\
&=& 
- {i\over 2\epsilon} {1\over M_{\rm pl}^2} {1\over k^3} 
\int_{-\infty}^{\eta'} {d\eta\over\eta^2}\, \bigl[ 
(1+k^2\eta\eta') \sin[k(\eta-\eta')]
-k(\eta-\eta') \cos[k(\eta-\eta')]
\bigr] 
\nonumber \\
&=& 
- {i\over 2\epsilon} {1\over M_{\rm pl}^2} {1\over k^3} 
\biggl[ -{\sin[k(\eta-\eta')]+k\eta'\cos[k(\eta-\eta')]\over\eta}
\biggr]_{-\infty}^{\eta'} 
= {i\over 2\epsilon} {1\over M_{\rm pl}^2} {1\over k^2} .
\nonumber 
\end{eqnarray}
Though it appeared to depend on $t'$ as well, none of this dependence survives in this leading result.  There remains just the $dt'$-integral,
\begin{eqnarray}
&&\!\!\!\!\!\!\!\!\!\!\!\!\!\!\!\!
i k^2 \int_{-\infty}^{t_*} dt\, 
{\dot\phi^2\over\dot\rho^2} e^{\rho(t)} \Bigl\{ 
G_2^{+++}(t,\vec k;t_*,\vec k_2;t_*,\vec k_3) 
- G_2^{-++}(t,\vec k;t_*,\vec k_2;t_*,\vec k_3) \Bigr\} 
\nonumber \\
&=& 
{i\beta\over 2\epsilon} 
{\dot\phi^2\over\dot\rho^2}\,\bigl[ k^2+k_2^2+k_3^2 \bigr]\, 
\int_{-\infty}^{t_*} dt'\, e^{\rho(t')} 
\bigl[ G_{k_2}^>(t_*,t')G_{k_3}^>(t_*,t') 
- G_{k_2}^<(t_*,t')G_{k_3}^<(t_*,t') \bigr]
\nonumber \\
&=& 
i\beta\, M_{\rm pl}^2\,\bigl[ k^2+k_2^2+k_3^2 \bigr]\, 
\int_{-\infty}^{t_*} dt'\, e^{\rho(t')} 
\bigl[ G_{k_2}^>(t_*,t')G_{k_3}^>(t_*,t') 
- G_{k_2}^<(t_*,t')G_{k_3}^<(t_*,t') \bigr] . 
\nonumber 
\end{eqnarray}
Here we have used $\dot\phi^2/\dot\rho^2=2\epsilon M_{\rm pl}^2$.  In performing this integral, we assume that $t_*$ is being taken in the late-time limit.  In terms of the conformal time, this limit corresponds to $\eta_*\to 0$, 
\begin{eqnarray}
&&\!\!\!\!\!\!\!\!\!\!\!\!\!\!\!\!
\int_{-\infty}^{t_*} dt'\, e^{\rho(t')} 
\bigl[ G_{k_2}^>(t_*,t')G_{k_3}^>(t_*,t') 
- G_{k_2}^<(t_*,t')G_{k_3}^<(t_*,t') \bigr] 
\nonumber \\
&=& 
{i\over 8\epsilon^2} {H^2\over M_{\rm pl}^4} 
{1\over k_2^3k_3^3} \int_{-\infty}^{\eta_*} {d\eta'\over\eta^{\prime\, 2}} \, 
\bigl[ 
(1-k_2k_3\eta^{\prime 2}) \sin[(k_2+k_3)\eta']
- (k_2+k_3)\eta' \cos[(k_2+k_3)\eta'] \bigr] 
\nonumber \\
&=& 
- {i\over 8\epsilon^2} {H^2\over M_{\rm pl}^4} 
{1\over k_2^3k_3^3} 
{k_2^2+k_3^2+k_2k_3\over k_2+k_3} + {\cal O}(\eta_*^2) . 
\nonumber 
\end{eqnarray}
Therefore, the leading contribution from the second operator to the consistency relation is 
\begin{eqnarray}
&&\!\!\!\!\!\!\!\!\!\!\!\!\!\!\!\!
i k^2 \int_{-\infty}^{t_*} dt\, 
{\dot\phi^2\over\dot\rho^2} e^{\rho(t)} \Bigl\{ 
G_2^{+++}(t,\vec k;t_*,\vec k_2;t_*,\vec k_3) 
- G_2^{-++}(t,\vec k;t_*,\vec k_2;t_*,\vec k_3) \Bigr\} 
\nonumber \\
&=& 
{\beta\over 8\epsilon^2} {H^2\over M_{\rm pl}^2} 
{k^2+k_2^2+k_3^2\over k_2^3k_3^3} 
{k_2^2+k_3^2+k_2k_3\over k_2+k_3} , 
\nonumber 
\end{eqnarray}
or rather, since it is meant to be evaluated in the limit where $k\to 0$ and $k_3=k_2$, 
$$
i k^2 \int_{-\infty}^{t_*} dt\, 
{\dot\phi^2\over\dot\rho^2} e^{\rho(t)} \Bigl\{ 
G_2^{+++}(t,\vec k;t_*,\vec k_2;t_*,\vec k_3) 
- G_2^{-++}(t,\vec k;t_*,\vec k_2;t_*,\vec k_3) \Bigr\} 
= {3\beta\over 8\epsilon^2} {H^2\over M_{\rm pl}^2} {1\over k_2^3} 
+ \cdots .
$$

Although we have evaluated this result in a slightly different language of Green's functions, the contribution from this particular operator in the cubic action is exactly the same as what appeared in Maldacena's calculation, once we have multiplied his expression in the appropriate `soft' limit by an inverse factor of the power spectrum.

\subsection{The first operator} 

While the combination of Wightman functions that appears in the contribution from the first operator to the difference of the three-point functions is more complicated than what we just encountered, there are many parallels between them.  To start, the operator ${\cal O}_1$'s contribution is 
\begin{eqnarray}
&&\!\!\!\!\!\!\!\!\!\!\!\!\!\!\!\!
G_1^{+++}(t,\vec k;t_*,\vec k_2;t_*,\vec k_3) 
- G_1^{-++}(t,\vec k;t_*,\vec k_2;t_*,\vec k_3) 
\nonumber \\
&=& 
-2i\alpha\, M_{\rm pl}^2\, \int_{-\infty}^{t_*} dt'\, e^{3\rho(t')} \Theta(t'-t)\, 
\nonumber \\
&&\qquad\qquad\Bigl\{
\bigl[ \dot G_k^>(t,t') - \dot G_k^<(t,t') \bigr]
\bigl[ \dot G_{k_2}^>(t_*,t')G_{k_3}^>(t_*,t') 
- \dot G_{k_2}^<(t_*,t')G_{k_3}^<(t_*,t') \bigr] 
\nonumber \\
&&\qquad\quad\ \ \,\,
+\, \bigl[ \dot G_k^>(t,t') - \dot G_k^<(t,t') \bigr]
\bigl[ G_{k_2}^>(t_*,t')\dot G_{k_3}^>(t_*,t') 
- G_{k_2}^<(t_*,t')\dot G_{k_3}^<(t_*,t') \bigr] 
\nonumber \\
&&\qquad\quad\ \ \,\,
+\, \bigl[ G_k^>(t,t') - G_k^<(t,t') \bigr]
\bigl[ \dot G_{k_2}^>(t_*,t')\dot G_{k_3}^>(t_*,t') 
- \dot G_{k_2}^<(t_*,t')\dot G_{k_3}^<(t_*,t') \bigr] \Bigr\} .
\nonumber 
\end{eqnarray}
The appearance of the $\Theta(t'-t)$ and the arrangement of the Wightman functions is exactly as before.  All that has changed is the appearance of the time-derivatives, which where inherited from the operator itself, ${\cal O}_1 = \alpha\, M_{\rm pl}^2\, e^{3\rho}\dot\zeta^2\zeta$, and which are permuted amongst the various coordinates.

The sole contribution to the consistency relation again comes from the spatial derivative, or $k^2$ part, of the 1PI Green's function acting on the difference of the three-point functions,
\begin{eqnarray}
&&\!\!\!\!\!\!\!\!\!\!\!\!\!\!\!\!
i k^2 \int_{-\infty}^{t_*} dt\, 
{\dot\phi^2\over\dot\rho^2} e^{\rho(t)} \Bigl\{ 
G_1^{+++}(t,\vec k;t_*,\vec k_2;t_*,\vec k_3) 
- G_1^{-++}(t,\vec k;t_*,\vec k_2;t_*,\vec k_3) \Bigr\} 
\nonumber \\
&=& 
2\alpha\, M_{\rm pl}^2\, k^2 {\dot\phi^2\over\dot\rho^2} 
\int_{-\infty}^{t_*} dt'\, e^{3\rho(t')} 
\biggl\{ \int_{-\infty}^{t'} dt\,  e^{\rho(t)}\, 
\bigl[ \dot G_k^>(t,t') - \dot G_k^<(t,t') \bigr] \biggr\}
\nonumber \\
&&\qquad\qquad\quad\quad\Bigl\{
\bigl[ \dot G_{k_2}^>(t_*,t')G_{k_3}^>(t_*,t') 
- G_{k_2}^<(t_*,t')G_{k_3}^<(t_*,t') \bigr] 
\nonumber \\
&&\qquad\qquad\quad\ \ \,\,
+\, \bigl[ G_{k_2}^>(t_*,t')\dot G_{k_3}^>(t_*,t') 
- G_{k_2}^<(t_*,t')\dot G_{k_3}^<(t_*,t') \bigr] 
 \Bigr\} 
\nonumber \\
&& 
+\,\, 2\alpha\, M_{\rm pl}^2\, k^2 {\dot\phi^2\over\dot\rho^2} 
\int_{-\infty}^{t_*} dt'\, e^{3\rho(t')} 
\bigl[ \dot G_{k_2}^>(t_*,t')\dot G_{k_3}^>(t_*,t') 
- \dot G_{k_2}^<(t_*,t')\dot G_{k_3}^<(t_*,t') \bigr] 
\nonumber \\
&&\qquad\qquad\quad\times
\int_{-\infty}^{t'} dt\, e^{\rho(t)}\, \bigl[ G_k^>(t,t')-G_k^<(t,t') \bigr] .
\nonumber 
\end{eqnarray}
One of these integrals---the integral over $t$---we have already encountered in treating ${\cal O}_2$, 
$$
\int_{-\infty}^{t'} dt\, e^{\rho(t)}\, \bigl[ G_k^>(t,t')-G_k^<(t,t') \bigr] 
= {i\over 2\epsilon} {1\over M_{\rm pl}^2} {1\over k^2} ,
$$
while the second of the integrals vanishes altogether,
\begin{eqnarray}
\int_{-\infty}^{t'} dt\, e^{\rho(t)} 
\bigl[ \dot G_k^>(t,t') - \dot G_k^<(t,t') \bigr] 
&=& 
{i\over 2\epsilon} {H\over M_{\rm pl}^2} {\eta^{\prime\, 2}\over k}
\int_{-\infty}^{\eta'} {d\eta\over\eta^2}\, 
\Bigl[ \sin[k(\eta-\eta')] - k\eta \cos[k(\eta-\eta')] \Bigr] 
\nonumber \\
&=& 
{i\over 2\epsilon} {H\over M_{\rm pl}^2} {\eta^{\prime\, 2}\over k}
\biggl[ -{\sin[k(\eta-\eta')]\over \eta} \biggr]_{-\infty}^{\eta'} 
= 0 .
\nonumber 
\end{eqnarray}
There remains one integral to perform, which we shall evaluate in the late-time limit ($\eta_*\to 0$) where $\eta_*$ essentially vanishes,
\begin{eqnarray}
&&\!\!\!\!\!\!\!\!\!\!\!\!\!\!\!\!\!\!\!\!\!\!\!\!\!\!\!\!\!\!\!\!
i k^2 \int_{-\infty}^{t_*} dt\, 
{\dot\phi^2\over\dot\rho^2} e^{\rho(t)} \Bigl\{ 
G_1^{+++}(t,\vec k;t_*,\vec k_2;t_*,\vec k_3) 
- G_1^{-++}(t,\vec k;t_*,\vec k_2;t_*,\vec k_3) \Bigr\} 
\nonumber \\
&=& 
2i\alpha\, M_{\rm pl}^2\,   
\int_{-\infty}^{t_*} dt'\, e^{3\rho(t')} 
\bigl[ \dot G_{k_2}^>(t_*,t')\dot G_{k_3}^>(t_*,t') 
- \dot G_{k_2}^<(t_*,t')\dot G_{k_3}^<(t_*,t') \bigr] 
\nonumber \\
&=& 
-\,\, {\alpha\over 4\epsilon^2} {H^2\over M_{\rm pl}^2} {1\over k_2k_3}
\int_{-\infty}^{\eta_*} d\eta'\, \sin[(k_2+k_3)\eta'] 
\nonumber \\
&=& 
{\alpha\over 4\epsilon^2} {H^2\over M_{\rm pl}^2} {1\over k_2k_3} 
{1\over k_2+k_3} + \cdots
\nonumber 
\end{eqnarray}

Going to the limit $k_3=k_2$ we have a genuine contribution to the left-side of the consistency relation from the first operator,
$$
\langle{\cal O}_1\rangle = 
{\alpha\over 8\epsilon^2} {H^2\over M_{\rm pl}^2} {1\over k_2^3} . 
$$

\subsection{The third operator} 

The third operator is the most complicated since it has both time and space derivatives, 
$$
{\cal O}_3 = \gamma\, M_{\rm pl}^2\, e^{3\rho} \dot\zeta \partial_k\zeta \partial^k(\partial^{-2}\dot\zeta) .
$$
It contributes the following to the difference of the three-point functions,
\begin{eqnarray}
&&\!\!\!\!\!\!\!
G_3^{+++}(t,\vec k;t_*,\vec k_2;t_*,\vec k_3) 
- G_3^{-++}(t,\vec k;t_*,\vec k_2;t_*,\vec k_3) 
\nonumber \\
&=& 
-i\gamma\, M_{\rm pl}^2\, \int_{-\infty}^{t_*} dt'\, e^{3\rho(t')} \Theta(t'-t)\, 
\nonumber \\
&&\quad\biggl\{
\biggl[ {\vec k\cdot\vec k_3\over k^2} + {\vec k_2\cdot\vec k_3\over k_2^2} \biggr]
\bigl[ \dot G_k^>(t,t') - \dot G_k^<(t,t') \bigr]
\bigl[ \dot G_{k_2}^>(t_*,t')G_{k_3}^>(t_*,t') 
- \dot G_{k_2}^<(t_*,t')G_{k_3}^<(t_*,t') \bigr] 
\nonumber \\
&&\quad
+\, \biggl[ {\vec k\cdot\vec k_2\over k^2} + {\vec k_3\cdot\vec k_2\over k_3^2} \biggr] 
\bigl[ \dot G_k^>(t,t') - \dot G_k^<(t,t') \bigr]
\bigl[ G_{k_2}^>(t_*,t')\dot G_{k_3}^>(t_*,t') 
- G_{k_2}^<(t_*,t')\dot G_{k_3}^<(t_*,t') \bigr] 
\nonumber \\
&&\quad
+\, \biggl[ {\vec k_2\cdot\vec k\over k_2^2} + {\vec k_3\cdot\vec k\over k_3^2} \biggr] 
\bigl[ G_k^>(t,t') - G_k^<(t,t') \bigr]
\bigl[ \dot G_{k_2}^>(t_*,t')\dot G_{k_3}^>(t_*,t') 
- \dot G_{k_2}^<(t_*,t')\dot G_{k_3}^<(t_*,t') \bigr] 
\biggr\} .
\nonumber 
\end{eqnarray}
The combinations of Wightman functions, although they are now accompanied by more complicated coefficients depending on the momenta, are nonetheless exactly what we saw while analysing ${\cal O}_1$.  So to analyse the contribution from ${\cal O}_3$ we do not need to reproduce each of the steps of the previous calculation.  

The contribution from the part due to the spatial derivatives is quickly evaluated by using results from the ${\cal O}_1$ calculation, 
\begin{eqnarray}
&&\!\!\!\!\!\!\!\!\!\!\!\!\!\!\!\!\!\!\!\!\!\!\!\!\!\!\!\!\!\!
i k^2 \int_{-\infty}^{t_*} dt\, 
{\dot\phi^2\over\dot\rho^2} e^{\rho(t)} \Bigl\{ 
G_3^{+++}(t,\vec k;t_*,\vec k_2;t_*,\vec k_3) 
- G_3^{-++}(t,\vec k;t_*,\vec k_2;t_*,\vec k_3) \Bigr\} 
\nonumber \\
&=& 
{\gamma\over 8\epsilon^2} {H^2\over M_{\rm pl}^2}
\biggl[ {\vec k_2\cdot\vec k\over k_2^2} + {\vec k_3\cdot\vec k\over k_3^2} \biggr]
{1\over k_2k_3}{1\over k_2+k_3} + \cdots .
\nonumber 
\end{eqnarray}
Notice that in the $\vec k\to\vec 0$ limit, this contribution goes away altogether.
$$
i k^2 \int_{-\infty}^{t_*} dt\, 
{\dot\phi^2\over\dot\rho^2} e^{\rho(t)} \Bigl\{ 
G_3^{+++}(t,\vec k;t_*,\vec k_2;t_*,\vec k_3) 
- G_3^{-++}(t,\vec k;t_*,\vec k_2;t_*,\vec k_3) \Bigr\} 
= -{\gamma\over 16\epsilon^2} {H^2\over M_{\rm pl}^2}
{k^2\over k_2^5} + \cdots = 0 .
$$
Therefore we have no contribution from this operator in the slow-roll, late-time, $k\to 0$ limits,
$$
\langle{\cal O}_3\rangle = 0 + \cdots . 
$$

\subsection{The left side of the consistency relation} 

If we add up what we have found, we find
$$
\langle{\cal O}_1\rangle + \langle{\cal O}_2\rangle + \langle{\cal O}_3\rangle
= {\alpha+3\beta\over 8\epsilon^2} {H^2\over M_{\rm pl}^2} {1\over k_2^3} .
$$
Putting in the values of $\alpha$, $\beta$, and $\gamma$, we have that
$$
\alpha + 3\beta 
= 4\epsilon [ 2\epsilon + \delta ] ,
$$
which yields the following leading result in the slow-roll parameters for the left side of the relation as we have written it, 
\begin{eqnarray}
&&\!\!\!\!\!\!\!\!\!\!\!\!\!\!\!\!\!\!\!\!
i \lim_{\vec k\to\vec 0} \int_{-\infty}^{t_*} dt\, 
\biggl\{ 
{\dot\phi^2\over\dot\rho^2} e^{3\rho} {d^2\over dt^2} 
+ {d\over dt}\biggl[ {\dot\phi^2\over\dot\rho^2} e^{3\rho} \biggr] {d\over dt}
+ k^2 {\dot\phi^2\over\dot\rho^2} e^\rho \biggr\}
\nonumber \\
&&\qquad\ 
\Bigl\{ 
G^{+++}_c(t,\vec k;t_*,\vec k_2;t_*,\vec k_3) 
- G^{-++}_c(t,\vec k;t_*,\vec k_2;t_*,\vec k_3) \Bigr\} 
\approx 
- {2\epsilon + \delta\over 2\epsilon} {H^2\over M_{\rm pl}^2} {1\over k_2^3}.
\nonumber 
\end{eqnarray}
The two-point function, or power spectrum, in the late time limit is 
$$
G_c^{++}(t_*,\vec k_2;t_*,-\vec k_2) \equiv P_{k_2}(t_*)
= {1\over 4\epsilon} {H^2\over M_{\rm pl}^2} 
{(-k_2\eta)^{-4\epsilon-2\delta}\over k_2^3} , 
$$
to leading order in the slow-roll parameters.  Taking the derivative that appears in the consistency relation, 
$$
- \bigl[ 3 + \vec k_2\cdot\nabla_{\vec k_2}\bigr]
G^{++}_c(t_*,\vec k_2;t_*,-\vec k_2) 
= - \biggl[ 3 + k_2 {\partial\over\partial k_2} \biggr]\, P_{k_2}(t_*)
= -{4\epsilon+2\delta\over 4\epsilon} {H^2\over M_{\rm pl}^2} {1\over k_2^3}
+ \cdots ,
$$
we do find that the two sides agree.

\section{Transforming the consistency relation into a standard form}

When we differentiated the Slavnov-Taylor identity with respect to the fields $\bar\zeta^{s_2}(y_2)$ and $\bar\zeta^{s_3}(y_3)$, we obtained  
\begin{eqnarray}
&&\!\!\!\!\!\!\!\!\!\!\!\!\!\!\!\!\!\!\!\!
\int d^4x\, \biggl\{ 
{\delta^2 J^+(x)\over\delta\bar\zeta^{s_2}(y_2)\delta\bar\zeta^{s_3}(y_3)} 
- {\delta^2 J^-(x)\over\delta\bar\zeta^{s_2}(y_2)\delta\bar\zeta^{s_3}(y_3)} 
\nonumber \\
&&\quad
+ \sum_{r,s=\pm} \int d^4z\, {\delta J^r(x)\over\delta\bar\zeta^{s_2}(y_2)}
{\delta J^s(z)\over\delta\bar\zeta^{s_3}(y_3)} \vec x\cdot\vec\nabla_{\vec x} {\delta W\over\delta J^r(x)\delta J^s(z)} 
\nonumber \\
&&\quad
+ \sum_{r,s=\pm} \int d^4z\, {\delta J^r(x)\over\delta\bar\zeta^{s_3}(y_3)}
{\delta J^s(z)\over\delta\bar\zeta^{s_2}(y_2)}
\vec x\cdot\vec\nabla_{\vec x} {\delta W\over\delta J^r(x)\delta J^s(z)} 
\biggr\} = 0 . 
\nonumber 
\end{eqnarray}
As a formal expression this is perfectly fine; but it is not yet in a form that is most suited for what is being measured.  It is the connected three-point function that is being constrained by observations rather than its one-particle irreducible counterpart,
$$
\Gamma^{\pm s_2s_3}(x,y_2,y_3) 
= {\delta^3\Gamma \over\delta\bar\zeta^\pm(x)\delta\bar\zeta^{s_2}(y_2)\delta\bar\zeta^{s_3}(y_3)} 
= \mp {\delta^2 J^\pm(x)\over\delta\bar\zeta^{s_2}(y_2)\delta\bar\zeta^{s_3}(y_3)} . 
$$
When we are not evolving from $t=-\infty$ to $\infty$, the higher-order 1PI Green's functions can be a bit difficult to calculate since we can no longer Fourier transform in all $3+1$ dimensions.  But just as in $S$-matrix calculations, the connected and 1PI three-point functions are related by amputating the propagators associated with the legs of the former to obtain the latter,
$$
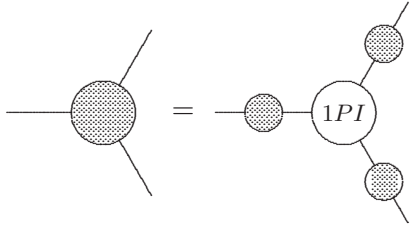
\beginpicture
\setcoordinatesystem units <1.00truept,1.00truept>
\setplotarea x from -36 to 18, y from -32 to 32
\circulararc 360 degrees from 12 0 center at 0 0
\plot -36 0  -12 0 /
\plot  6  10.392  18  31.177 /
\plot  6 -10.392  18 -31.177 /
\setshadesymbol ({\tmrms .})
\setshadegrid span <0.9pt>
\setquadratic
\hshade -8.475 -8.475 -8.475 <,z,,>  0 -12.0   -8.475 8.475 -8.475 -8.475 /
\vshade -8.475 -8.475  8.475 <z,z,,> 0 -12.0   12.0   8.475 -8.475  8.475 /
\hshade -8.475  8.475  8.475 <z,,,>  0  8.475  12.0   8.475  8.475  8.475 /
\endpicture
\beginpicture
\setcoordinatesystem units <1.00truept,1.00truept>
\setplotarea x from -72 to 48, y from -42 to 42
\circulararc 360 degrees from 12 0 center at 0 0
\circulararc 360 degrees from -37 0 center at -30 0
\circulararc 360 degrees from 11.5  19.919 center at 15  25.981
\circulararc 360 degrees from 11.5 -19.919 center at 15 -25.981
\plot -48 0  -37 0 /
\plot -23 0  -12 0 /
\plot 6  10.392  11.5  19.919 /
\plot 18.5  32.043  24  41.569 /
\plot 6 -10.392  11.5 -19.919 /
\plot 18.5 -32.043  24 -41.569 /
\put {$=$} [r] at -56 0
\put {{\footnotesize $1PI$}} [c] at 0 0
\setshadesymbol ({\tmrms .})
\setshadegrid span <0.9pt>
\setquadratic
\hshade  -4.94 -34.94 -34.94 <,z,,>    0 -37.0  -34.94  4.94 -34.94 -34.94 /
\vshade -34.94  -4.94   4.94 <z,z,,> -30  -7.0    7.0 -25.06  -4.94   4.94 /
\hshade  -4.94 -25.06 -25.06 <z,,,>    0 -25.06 -23.0   4.94 -25.06 -25.06 /
\hshade 21.04 10.06 10.06 <,z,,>  25.98  8.0   10.06  30.92 10.06 10.06 /
\vshade 10.06 21.04 30.92 <z,z,,> 15    18.98  32.98  19.94 21.04 30.92 /
\hshade 21.04 19.94 19.94 <z,,,>  25.98 19.94  22.0   30.92 19.94 19.94 /
\hshade -21.04  10.06  10.06 <,z,,>  -25.98  8.0   10.06 -30.92  10.06  10.06 /
\vshade  10.06 -30.92 -21.04 <z,z,,>  15   -32.98 -18.98  19.94 -30.92 -21.04 /
\hshade -21.04  19.94  19.94 <z,,,>  -25.98 19.94  22.0  -30.92  19.94  19.94 /
%
\endpicture
$$
The only extra complication is that unlike the analogous $S$-matrix statement we must sum over the $\pm$ indices in addition to integrating over the intermediate positions where the propagators are attached.  By applying the appropriate operator, we can convert the 1PI three-point functions in our expression into connected three-point functions on which the 1PI two-point functions are acting.  

Let us apply the following operator to the previous version of the relation,
$$
\sum_{s_2,s_3=\pm}\int d^4y_2\, \int d^4y_3\, 
{\delta\bar\zeta^{s_2}(y_2)\over\delta J^+(x_2)}
{\delta\bar\zeta^{s_3}(y_3)\over\delta J^+(x_3)} . 
$$
If we think of the consistency relation as a relation amongst different graphs, what this operator does is to attach connected two-point functions (propagators) to two of the external legs of each term in the relation.  Analytically, the relation becomes
\begin{eqnarray}
&&\!\!\!\!\!\!\!\!\!\!\!\!\!\!\!\!\!\!\!\!
\sum_{s_2,s_3=\pm} \int d^4y\, \int d^4y_2\, \int d^4y_3\, 
{\delta\bar\zeta^{s_2}(y_2)\over\delta J^+(x_2)}
{\delta\bar\zeta^{s_3}(y_3)\over\delta J^+(x_3)} 
\biggl\{ 
{\delta^2 J^+(y)\over\delta\bar\zeta^{s_2}(y_2)\delta\bar\zeta^{s_3}(y_3)} 
- {\delta^2 J^-(y)\over\delta\bar\zeta^{s_2}(y_2)\delta\bar\zeta^{s_3}(y_3)} 
\biggr\} 
\nonumber \\
&&\qquad\qquad\qquad\qquad\qquad\qquad\qquad\qquad
= - \bigl[ \vec x_2\cdot\vec\nabla_{\vec x_2} 
+ \vec x_3\cdot\vec\nabla_{\vec x_3} \bigr] 
{\delta W\over\delta J^+(x_2)\delta J^+(x_3)} 
\nonumber 
\end{eqnarray}
For later convenience, we have switched one of the integration variables from $x$ to $y$.  Notice that one of the effects of acting with this operator has been to remove the 1PI two-point functions entirely from the terms with the directional derivatives, $\vec x\cdot\vec\nabla_{\vec x}$.  This happened because they are inverses of the functional derivatives in the operator that we applied,
$$
\sum_{s=\pm}\int d^4z\, 
{\delta\bar\zeta^s(z)\over\delta J^{r_1}(x)}
{\delta J^{r_2}(y)\over\delta\bar\zeta^s(z)} 
= \delta^{r_2}_{r_1}\, \delta^4(x-y) . 
$$
If we differentiate this identity with respect to a source $J^+$, and set $r=+$ and $s=s_2$, we have a way of rewriting the left side of the consistency relation in terms of connected three-point functions rather than the 1PI ones, 
\begin{eqnarray}
&&\!\!\!\!\!\!\!\!\!\!\!\!\!\!\!\!\!\!\!\!
\sum_{s_2,s_3}\int d^4y_2\, \int d^4y_3\, 
\biggl\{ 
{\delta\bar\zeta^{s_2}(y_2)\over\delta J^+(x_2)}
{\delta\bar\zeta^{s_3}(y_3)\over\delta J^+(x_3)} 
{\delta^2 J^r(y)\over\delta\bar\zeta^{s_2}(y_2)\delta\bar\zeta^{s_3}(y_3)}
\biggr\}
\nonumber \\
&&\qquad\qquad\qquad\qquad\qquad\qquad =  
- \sum_{s=\pm} \int d^4x\, \biggl\{ 
{\delta J^r(y)\over\delta\bar\zeta^s(x)}
{\delta^2 \bar\zeta^s(x)\over\delta J^+(x_2)\delta J^+(x_3)}
\biggr\} . 
\nonumber 
\end{eqnarray}
The consistency relation then assumes the form that we have been seeking,
\begin{eqnarray}
&&\!\!\!\!\!\!\!\!\!\!\!\!\!\!\!\!\!\!\!\!
\sum_{s=\pm} \int d^4y\, \int d^4x\, 
\biggl\{ 
{\delta J^+(y)\over\delta\bar\zeta^s(x)}
{\delta^2 \bar\zeta^s(x)\over\delta J^+(x_2)\delta J^+(x_3)}
- {\delta J^-(y)\over\delta\bar\zeta^s(x)}
{\delta^2 \bar\zeta^s(x)\over\delta J^+(x_2)\delta J^+(x_3)}
\biggr\} 
\nonumber \\
&&\qquad\qquad\qquad\qquad\qquad\qquad\qquad\qquad
= \bigl[ \vec x_2\cdot\vec\nabla_{\vec x_2} 
+ \vec x_3\cdot\vec\nabla_{\vec x_3} \bigr] 
{\delta W\over\delta J^+(x_2)\delta J^+(x_3)} . 
\nonumber 
\end{eqnarray}

All that is left is to convert the functions that appear in this relation into a more conventional form, using the definitions of the connected and 1PI Green's functions in terms of functional derivatives of the appropriate generating functional, 
\begin{eqnarray}
&&\!\!\!\!\!\!\!\!\!\!\!\!\!\!\!\!\!\!\!\!
-i\sum_{s=\pm}\int d^4y\, \int d^4x\, 
\Bigl\{ \Gamma^{+s}(y,x)\, G^{s++}_c(x,x_2,x_3) 
+ \Gamma^{-s}(y,x)\, G^{s++}_c(x,x_2,x_3) \Bigr\} 
\nonumber \\
&&\qquad\qquad\qquad\qquad\qquad\qquad\qquad
= \bigl[ \vec x_2\cdot\vec\nabla_{\vec x_2} 
+ \vec x_3\cdot\vec\nabla_{\vec x_3} \bigr] G^{++}_c(x_2,x_3) . 
\nonumber 
\end{eqnarray}

\end{document}